\documentclass[prd,aps,showkeys,floats,onecolumn]{revtex4}
\usepackage{amsmath}
\usepackage{amssymb}

\def\semidirect{\;{\rlap{$\subset$}\times}\;}

\def\sqr#1#2{{\vcenter{\hrule height.#2pt\hbox{\vrule width.#2pt
height#1pt \kern#1pt \vrule width.#2pt}\hrule height.#2pt}}}
\def\square{\mathchoice\sqr64\sqr64\sqr{4.2}3\sqr{3.0}3}

\begin{document}

\title[TSSD induced dynamical torsion]{Two-step spacetime deformation induced dynamical torsion}

\author{G Ter-Kazarian}

\address{Byurakan Astrophysical Observatory,
Byurakan 378433, Aragatsotn District, Armenia, E-mail:
gago\_50@yahoo.com}

\begin{abstract}
We extend the geometrical ideas of the spacetime deformations to
study the physical foundation of the post-Riemannian geometry. To
this aim, we construct the theory of {\em two-step spacetime
deformation} as a guiding principle. We address the theory of
teleparallel gravity and construct a consistent Einstein-Cartan (EC)
theory with the {\em dynamical torsion}. We show that the equations
of the standard EC theory, in which the equation defining torsion is
the algebraic type and, in fact, no propagation of torsion is
allowed, can be equivalently replaced by the set of {\em modified EC
equations} in which the torsion, in general, is  {\em dynamical}.
The special physical constraint imposed upon the spacetime
deformations yields the short-range propagating spin-spin
interaction.
\end{abstract}

\keywords{Spacetime deformations, Post-Riemannian geometry,
Dynamical torsion}

\maketitle

\section{Introduction}
At present, the papers on the gauge treatment of gravity are the
considerable part of all gravitational investigations. In its
present formulation, this exploits  the language of the fundamental
geometric structure known as a fiber bundle, which provides a
unified picture of gravity modified models based on several Lie
groups, see e.g.~\cite{rf2}-\cite{rf13}. These efforts mainly focus
on the physical foundation of torsion, and its connection to quantum
gravity and microphysics. The Einstein-Cartan (EC) theory is the
minimal extension of the general relativity, which considers
curvature and torsion as representing independent degrees of
freedom, and relating torsion to the density of intrinsic angular
momentum. In the standard EC theory, the equation defining torsion
is of algebraic type, and not a differential equation, and that no
propagation of torsion is allowed. As is known from the weak
interaction, the causality reasons do not respect a contact
interaction. Therefore, many modifications of the EC theory have
been proposed in recent years, see e.g.~\cite{rf10}, right up to
introducing a higher order gravitational Lagrangian of quadratic
models~\cite{rf14}, different from the simple scalar curvature. This
aimed to obtain a differential field equation for the torsion
tensor, instead of an algebraic one. Even though a strong emphasis
has been placed on this issue throughout the development of modern
physics, all these approaches are subject to many uncertainties. On
the other hand,  a general way to deform the spacetime metric with
constant curvature has been explicitly posed by
~\cite{rf15}-\cite{rf17}. The problem was initially solved only for
3D spaces, but consequently it was solved also for spacetimes of any
dimension. It was proved that any semi-Riemannian metric can be
obtained as a deformation of constant curvature matric, this
deformation being parameterized by a 2-form. These results are fully
recovered and generalized by ~\cite{rf11}, where a novel definition
of spacetime metric deformations, parameterized in terms of scalar
field matrices, is proposed.

In this article we construct the theory of {\em two-step spacetime
deformation} (TSSD), which generalizes and, in particular cases,
fully reproduces  the results of the conventional theory of
spacetime deformation. We show that through a non-trivial choice of
explicit form of a {\em deformation} tensor, we have a way to derive
different post Riemannian spacetime structures such as:\phantom{a}
1) the Weitzenb\"{o}ck space, $W_{4}$, underlying a teleparallelism
theory of gravity, see e.g.~\cite{rf18}; \phantom{a} 2) the RC
manifold, $U_{4}$, underlying Einstein-Cartan theory also called
Einstein-Cartan-Sciama-Kibble, for a comprehensive references, see
for example~\cite{rf9,rf10,rf13}); \phantom{a} 3) or even the most
general linear connection of MAG theory taking values in the
Lie-algebra of the 4D-affine group, $A(4, R) =R^4\semidirect
GL(4,R)$ (the semi-direct product of the group of 4D-translations
and general linear 4D-transformations), see e.g.~\cite{rf7,rf8}. We
are mainly interested in the formulation of physical aspects of the
EC theory from a novel, TSSD, view point. In the framework of the
TSSD-$U_{4}$ theory, we address the key problem of the {\em
dynamical torsion}. We show that the equations of the standard EC
theory can be equivalently replaced by the set of {\em modified EC
equations} in which the torsion, in general, is a {\em dynamical}.
The physical constraint will be imposed upon the spacetime
deformations, which yields the short-range propagating spin-spin
interaction. We will proceed according to the following structure.
In section 2 we construct the TSSD theory as the guiding principle.
In section 3, in case of particular spacetime deformations, we
obtain the theory of teleparallel gravity. In section 4, emerging
structures are embedded into the foundations of the TSSD- $U_{4}$
theory  in both a tensorial form and a language of the differential
forms, and the equation of the short-range propagating torsion is
derived. The concluding remarks are presented in section 5.  We use
the Greek alphabet ($\mu, \nu, \rho,...= 0, 1, 2, 3$) to denote the
holonomic world indices related to curved spacetime
$\mathcal{M}_{4}$, and the Latin alphabet ($a, b, c, . . . = 0, 1,
2, 3$) to denote the anholonomic indices related to the tangent
space.

\section{The TSSD as a guiding principle} When considering several connections with
different curvature and torsion, one takes spacetime simply as a
manifold, and connections as additional structures, see
e.g.~\cite{rf20,rf21}. From this view point, below we shall tackle
the problem of spacetime deformation. To start with, let us consider
the holonomic metric defined in the Riemann space, $V_{4}$, as
\begin{equation}
\begin{array}{l}
\breve{g}=\breve{g}_{\mu\nu}\,\breve{\vartheta}{}^{\mu}\otimes\breve{\vartheta}{}^{\nu}=
\breve{g}(\breve{e}_{\mu}, \,\breve{e}_{\nu})\,
\breve{\vartheta}{}^{\mu}\otimes\breve{\vartheta}{}^{\nu},
\end{array}
\label{R24}
\end{equation}
with components, $\breve{g}_{\mu\nu}=\breve{g}(\breve{e}_{\mu},
\breve{e}_{\nu})$ in the dual holonomic base
$\{\breve{\vartheta}{}^{\mu}\equiv d\breve{x}{}^{\mu}\}$. All
magnitudes related to the Riemann space, $V_{4}$, will be denoted
with an over ${}^{\prime}\phantom{a}\breve{}\phantom{a}{}^{\prime}$.
The space, $V_{4}$, has at each point a tangent space,
$\breve{T}_{\breve{x}}V_{4}$, spanned by the four tetrad fields,
$\breve{e}_{a}=\breve{e}_{a}^{\phantom{a}\mu}\,\breve{\partial}_{\mu}$,
which relate $\breve{g}$ to the tangent space metric, $o_{ab} =
diag(+ - - -)$, by
\begin{equation}
\begin{array}{l}
o_{ab} = \breve{g}(\breve{e}_{a}, \,\breve{e}_{b})=
\breve{g}_{\mu\nu}\,\breve{e}_{a}^{\phantom{a}\mu}\,\breve{e}_{b}^{\phantom{a}\nu}.
\end{array}
\label{RF0}
\end{equation}
The coframe members are
$\breve{\vartheta}{}^{b}=\breve{e}{}^{b}_{\phantom{a}
\mu}\,d\breve{x}{}^{\mu}$, such that $\breve{e}_{a}\,\rfloor\,
\breve{\vartheta}{}^{b}=\delta^{b}_{a}$,  where $\rfloor$ denoting
the interior product, namely, this is a $C^{\infty}$-bilinear map
$\rfloor:\Omega^{1}\rightarrow \Omega^{0}$ with $\Omega^{p}$ denotes
the $C^{\infty}$-modulo of differential p-forms on $V_{4}$. In
components
$\breve{e}_{a}^{\phantom{a}\mu}\,\breve{e}{}^{b}_{\phantom{a}\mu}=\delta^{b}_{a}$.
One can consider general transformations of the linear group, $GL(4,
R)$, taking any base into any other set of four linearly independent
fields. The notation, $\{\breve{e}_{a},\,\breve{\vartheta}{}^{b}\}$,
will be used below for general linear frames. Relation~(\ref{RF0})
has the converse
$\breve{g}_{\mu\nu}=o_{ab}\,\breve{e}{}^{a}_{\phantom{a}\mu}\,\breve{e}{}^{b}_{\phantom{a}\nu}$
because
$\breve{e}_{a}^{\phantom{a}\mu}\,\breve{e}{}^{a}_{\phantom{a}\nu}=\delta^{\mu}_{\nu}$.
The anholonomy objects read
\begin{equation}
\begin{array}{l}
\breve{C}{}^{a}:\phantom{a}=d\,\breve{\vartheta}{}^{a}=
\frac{1}{2}\,\breve{C}{}_{\phantom{a}bc}^{a}\,\breve{\vartheta}{}^{b}\,\wedge\,
\breve{\vartheta}{}^{c},
\end{array}
\label{R44}
\end{equation}
where the anholonomy coefficients,
$\breve{C}{}_{\phantom{a}bc}^{a}$, which represent the curls of the
base members are
\begin{equation}
\begin{array}{l}
\breve{C}{}_{\phantom{a}ab}^{c}=-\breve{\vartheta}{}^{c}([\breve{e}_{a},\,
\breve{e}_{b}])=
\breve{e}_{a}^{\phantom{a}\mu}\breve{e}_{b}^{\phantom{a}\nu}(\breve{\partial}_{\mu}\breve{e}{}^{c}_{\phantom{a}\nu}-
\breve{\partial}_{\nu}\breve{e}{}^{c}_{\phantom{a}\mu})=-\breve{e}{}^{c}_{\phantom{a}\mu}[\breve{e}_{a}(\breve{e}_{b}^{\phantom{a}\mu})-
\breve{e}_{b}(\breve{e}_{a}^{\phantom{a}\mu})].
\end{array}
\label{R26A2}
\end{equation}
The (anholonomic) Levi-Civita (or Christoffel) connection can be
written as
\begin{equation}
\begin{array}{l}
\breve{\Gamma}{}_{ab}:\phantom{a}= \breve{e}_{[a}\rfloor
d\breve{\vartheta}_{b]} - {1\over 2}\,(\breve{e}_{a}\rfloor
\breve{e}_{b}\rfloor
d\breve{\vartheta}_{c})\,\wedge\,\breve{\vartheta}{}^{c}
\end{array}
\label{R27T}
\end{equation}
where $\breve{\vartheta}_{c}$ is understood as the down indexed
1-form $\breve{\vartheta}_c=o_{cb}\,\breve{\vartheta}{}^b$.

\subsection{Model building:\phantom{a} spacetime deformation}
Next, we write the norm, $ds$, of the infinitesimal displacement,
$d\,x^{\mu}$, on the general smooth differential 4D-manifold
$\mathcal{M}_{4}$, in terms of the spacetime structures of $V_{4}$,
as
\begin{equation}
\begin{array}{l}
ds=\Omega_{\mu}^{\phantom{a}\nu}\breve{e}_{\nu}
\breve{\vartheta}{}^{\mu}=\Omega_{b}^{\phantom{a}a}\,\breve{e}_{a}\,\breve{\vartheta}{}^{b}=
e_{\rho}\,\vartheta^{\rho}=e_{a}\,\vartheta^{a}\,\in\,\mathcal{M}_{4},
\end{array}
\label{R28}
\end{equation}
where $\Omega_{\mu}^{\phantom{a}\nu}$ is the world-{\em deformation
tensor}, $\{e_{a}=e_{a}^{\phantom{a}\rho}\,e_{\rho}\}$ is the frame
and $\{\vartheta^{a}=e^{a}_{\phantom{a}\rho}\,\vartheta^{\rho}\}$ is
the coframe defined on $\mathcal{M}_{4}$, such that
$e_{a}\,\rfloor\, \vartheta^{b}=\delta^{b}_{a}$, or in components,
$e_{a}^{\phantom{a}\mu}\, e^{b}_{\phantom{a}\mu}=\delta^{b}_{a}$,
also the procedure can be inverted
$e^{a}_{\phantom{a}\rho}\,e_{a}^{\phantom{a}
\sigma}=\delta^{\sigma}_{\rho}$, provided
\begin{equation}
\begin{array}{l}
\Omega^{\phantom{a}\nu}_{\mu}=\pi^{\phantom{a}\rho}_{\mu}\,\pi_{\rho}^{\phantom{a}\nu},\quad
\Omega^{\phantom{a}a}_{b}=\pi^{\phantom{a}a}_{c}\,\pi^{\phantom{a}c}_{b}=
\Omega^{\phantom{a}\nu}_{\mu}\,\breve{e}{}^{a}_{\phantom{a}\nu}\,\breve{e}_{b}^{\phantom{a}\mu},\quad
e_{\rho}=\pi_{\rho}^{\phantom{a}\nu}\,\breve{e}_{\nu}\equiv
\partial_{\rho},\\

\vartheta^{\rho}=\pi^{\rho}_{\phantom{a}\mu}\,\breve{\vartheta}{}^{\mu}\equiv
d\,x^{\rho},\quad x^{\rho}\,\in\, {\cal U}\in\mathcal{M}_{4}.
\end{array}
\label{R29}
\end{equation}
Hence the deformation tensor, $\Omega^{a}_{\phantom{a}b}$, yields
local tetrad deformations
\begin{equation}
\begin{array}{l}
e_{a}\vartheta^{a}=\Omega^{a}_{\phantom{a}b}\,\breve{e}_{a}\,\breve{\vartheta}{}^{b},\quad
e_{c}=\pi^{\phantom{a}a}_{c}\,\breve{e}_{a},\quad
\vartheta^{c}=\pi_{\phantom{a}b}^{c}\,\breve{\vartheta}{}^{b}.
\end{array}
\label{R30}
\end{equation}
A general spin connection then transforms according
\begin{equation}
\begin{array}{l}
\omega^{a}_{\phantom{a}b\mu}=\pi^{\phantom{a}a}_{c}
\breve{\omega}{}^{c}_{\phantom{a}d\mu} \pi_{\phantom{b}b}^{d}+
\pi^{\phantom{a}a}_{c}\, \partial_{\mu} \,\pi_{\phantom{b}b}^{c}.
\end{array}
\label{R30G}
\end{equation}
Deformations~(\ref{R30}) restore a formalism of the spacetime metric
deformation proposed in \cite{rf11}. Therefore, following this work,
the matrices, $\pi(x):\phantom{a}=(\pi^{\phantom{a}a}_{b})(x)$, can
be called {\em first deformation matrices}, and the matrices
\begin{equation}
\begin{array}{l}
\gamma_{cd}(x)=o_{ab}\,\pi^{\phantom{a}a}_{c}(x)\,\pi_{d}^{\phantom{a}b}(x),
\end{array}
\label{R31}
\end{equation}
{\em second deformation matrices}. The matrices,  $
\pi_{\phantom{a}c}^{a}(x)\,\in\, GL(4, R)\,\forall\, x, $ in
general, give rise to the right cosets of the Lorentz group, i.e.
they are the elements of the quotient group $GL(4, R)/SO(3,1)$. If
we deform the tetrad according to (\ref{R30}), in general, we have
two choices to recast metric as follows: either writing the
deformation of the metric in the space of tetrads or deforming the
tetrad field:
\begin{equation}
\begin{array}{l}
g=o_{ab}\,\pi_{\phantom{a}c}^{a}\pi_{\,\,\,d}^{b}\breve{\vartheta}{}^{c}\otimes
\breve{\vartheta}{}^{d}= \gamma_{cd}\,\breve{\vartheta}{}^{c}\otimes
\breve{\vartheta}{}^{d}=o_{ab}\,\vartheta^{a}\otimes \vartheta^{b}.
\end{array}
\label{R34}
\end{equation}
In the first case, the contribution of the Christoffel symbols,
constructed by the metric $\gamma_{ab}$, reads
\begin{equation}
\begin{array}{l}
\Gamma_{\phantom{a}bc}^{a}=\frac{1}{2}\,\left(\breve{C}{}_{\phantom{a}bc}^{a}-\gamma^{aa'}\,\gamma_{bb'}\,\breve{C}{}_{\phantom{a}a'c}^{b'}-
    \gamma^{aa'}\,\gamma_{cc'}\,\breve{C}{}_{\phantom{a}a'b }^{c'} \right)\\
+ \frac{1}{2}\,\gamma^{aa'}\,\left(\breve{e}_{c} \,\rfloor\,
d\,\gamma_{ba'}- \breve{e}_{b}\,\rfloor\, d\,\gamma_{ca'}
     - \breve{e}_{a'} \,\rfloor \,d\,\gamma_{bc}\right).
\end{array}
\label{R35}
\end{equation}
The second deformation matrix, $\gamma_{ab}$, can be decomposed in
terms of symmetric, $\pi_{(ab)}$, and antisymmetric, $\pi_{[ab]}$,
parts of the matrix $\pi_{ab}=o_{ac}\pi^{c}_{\phantom{a}b}$ as
\begin{equation}
\begin{array}{l}
\gamma_{ab}=\Upsilon^{2}\,o_{ab}+2\Upsilon\,\Theta_{ab}+o_{cd}\,\Theta^{c}_{\phantom{a}a}\,\Theta^{d}_{\phantom{a}b}+
o_{cd}\,(\Theta^{c}_{\phantom{a}a}\,\varphi^{d}_{\phantom{a}b}+
\varphi^{c}_{\phantom{a}a}\,\Theta^{d}_{\phantom{a}b})+o_{cd}\,\varphi^{c}_{\phantom{a}a}\,\varphi^{d}_{\phantom{a}b},
\end{array}
\label{R36}
\end{equation}
where
\begin{equation}
\begin{array}{l}
\pi_{ab}=\Upsilon o_{ab}+\Theta_{ab}+\varphi_{ab}
\end{array}
\label{R37}
\end{equation}
$\Upsilon=\pi^{a}_{a}$, $\,\,\Theta_{ab}$ is the traceless symmetric
part and $\varphi_{ab}$ is the skew symmetric part of the first
deformation matrix. Consequently, the deformed metric, can be split
as
\begin{equation}
\begin{array}{l}
g_{\mu\nu}(\pi)=\Upsilon^{2}(\pi)\,\breve{g}_{\mu\nu}+\gamma_{\mu\nu}(\pi),
\end{array}
\label{R38}
\end{equation}
where
\begin{equation}
\begin{array}{l}
\gamma_{\mu\nu}(\pi)=[\gamma_{ab}-\Upsilon^{2}(\pi)\,o_{ab}]\,
\breve{e}{}^{a}_{\phantom{a}\mu}\,\breve{e}{}^{b}_{\phantom{a}\nu}.
\end{array}
\label{R39}
\end{equation}
The inverse deformed metric reads
\begin{equation}
\begin{array}{l}
g^{\mu\nu}(\pi)=o^{cd}\,\pi^{-1}{}_{\phantom{a}c}^{a}\,\pi^{-1}{}_{\phantom{a}d}^{b}\,\breve{e}_{a}^{\phantom{a}\mu}\,\breve{e}_{b}^{\phantom{a}\nu},
\end{array}
\label{R33}
\end{equation}
where
$\pi^{-1}{}_{\phantom{a}c}^{a}\,\pi_{\phantom{a}b}^{c}=\pi_{\phantom{a}b}^{c}\,\pi^{-1}{}_{\phantom{a}c}^{a}=\delta^{a}_{b}.$
In the second case, let us write the commutation table for the
anholonomic frame, $\{e_{a}\}$,
\begin{equation}
\begin{array}{l}
[e_{a},\, e_{b}]=- \frac{1}{2}\,C_{\phantom{a}ab}^{c}\,e_{c},
\end{array}
\label{R26A3}
\end{equation}
and define a dual expression of the new anholonomy objects,
$C_{\phantom{a}bc}^{a}$,
\begin{equation}
\begin{array}{l}
C^{a}:\phantom{a}=d\,\vartheta^{a}=
\frac{1}{2}\,C_{\phantom{a}bc}^{a}\,\vartheta^{b}\,\wedge\,
\vartheta^{c}=\frac{1}{2}\,(\partial_{\mu}\,e^{c}_{\phantom{a}\nu}-
\partial_{\nu}\,e^{c}_{\phantom{a}\mu})\,d\,x{}^{\mu}\,\wedge\,
d\,x{}^{\nu},
\end{array}
\label{R44}
\end{equation}
where
\begin{equation}
\begin{array}{l}
C_{\phantom{a}bc}^{a}=\pi^{a}_{\phantom{a}e}\,
{\pi^{-1}}_{\phantom{d}b}^{d} \,{\pi^{-1}}_{\phantom{f}c}^{f}\,
\breve{C}{}_{\,df}^{e}+2\,\pi^{a}_{\phantom{a}f}\,\breve{e}_{g}^{\phantom{a}\mu}
\,\left({\pi^{-1}}_{\phantom{g}\,[b}^{g}\partial{}_{\mu}\,{\pi^{-1}}_{\phantom{f}c]}^{f}\right).
\end{array}
\label{R45}
\end{equation}
In the particular case of constant metric in  tetradic space, the
deformed connection can be written as
\begin{equation}
\begin{array}{l}
\Gamma_{\phantom{a}bc}^{a}=\frac{1}{2}\,\left(C_{\phantom{a}bc}^{a}-o^{aa'}\,o_{bb'}\,C_{\phantom{a}a'c}^{b'}-
    o^{aa'}\,o_{cc'}\,C_{\phantom{a}a'b }^{c'} \right).
\end{array}
\label{R46}
\end{equation}
The usual Levi-Civita connection corresponding to the metric
(\ref{R34}) is related to the original connection by the relation
\begin{equation}
\begin{array}{l}
\Gamma^{\mu}_{\phantom{a}\rho\sigma}=\breve{\Gamma}{}^{\mu}_{\phantom{a}\rho\sigma}+\Pi^{\mu}_{\phantom{a}\rho\sigma},
\end{array}
\label{R40}
\end{equation}
provided
\begin{equation}
\begin{array}{l}
\Pi^{\mu}_{\phantom{a}\rho\sigma}=2g^{\mu\nu}\,\breve{g}_{\nu(\rho}\,\nabla_{\sigma)}\,\Upsilon-
\breve{g}_{\rho\sigma}\,g^{\mu\nu}\,\nabla_{\nu}\,\Upsilon+
\frac{1}{2}\,g^{\mu\nu}\, (\nabla_{\rho}\,\gamma_{\nu\sigma}+
\nabla_{\sigma}\,\gamma_{\rho\nu}-\nabla_{\nu}\,\gamma_{\rho\sigma}),
\end{array}
\label{R41}
\end{equation}
where the controvariant deformed metric, $g^{\nu\rho}$, is defined
as the inverse of $g_{\mu\nu}$, such that
$g_{\mu\nu}\,g^{\nu\rho}=\delta^{\rho}_{\mu}$. Hence, the connection
deformation $\Pi^{\mu}_{\phantom{a}\rho\sigma}$ acts like a force
that deviates the test particles from the geodesic motion in the
space, $V_{4}$.

\subsection{The post-Riemannian geometry}
We now assume that a deformation
$(\breve{e},\,\breve{\vartheta})\rightarrow (e,\,\vartheta)$ is
performed, according to the following heuristic map, in {\em
two-steps}:
\begin{center}
\begin{picture}(70,80)

\put(7,57){\vector(1,-1){30}}

\put(35,25){\vector(-1,-1){30}}

\put(2,57){\vector(0,-5){62}}

\put(-7,65){$(\breve{e}(\breve{x}) ,\,\breve{\vartheta}(\breve{x})
)$}

\put(-7,-15){$(e(x),\,\vartheta(x))$}

\put(45,25){$(\stackrel{\bullet}{e}(\stackrel{\bullet}{x}),\,\stackrel{\bullet}{\vartheta}(\stackrel{\bullet}{x}))$}

\put(-23,20){$\pi(x)$}

\put(27,45){$\stackrel{\bullet}{\pi}(\stackrel{\bullet}{x})$}

\put(25,4){$\sigma(x)$}

\put(-50,-35){Two-step deformation map}

\end{picture}
\end{center}
\vskip 1.5truecm  provided, we require that the first deformation
matrix,
$\stackrel{\bullet}{\pi}(\stackrel{\bullet}{x}):\phantom{a}=(\stackrel{\bullet}{\pi}{}^{\phantom{a}a}_{b})(\stackrel{\bullet}{x})$,
satisfies the following peculiar condition:
\begin{equation}
\begin{array}{l}
\stackrel{\bullet}{\pi}{}^{\phantom{a}a}_{c}(\stackrel{\bullet}{x})\,\stackrel{\bullet}{\partial}_{\mu}
\,\stackrel{\bullet}{\pi^{-1}}{}_{\phantom{b}b}^{c}(\stackrel{\bullet}{x})=\breve{\omega}{}^{a}_{\phantom{a}b\mu}(\breve{x}),
\end{array}
\label{R49}
\end{equation}
where $\breve{\omega}{}^{a}_{\phantom{a}b\mu}(\breve{x})$ is the
spin connection defined in the Riemann space. Whereas,
\begin{equation}
\begin{array}{l}
\stackrel{\bullet}{\Omega}{}^{\phantom{a}\nu}_{\mu}=
\stackrel{\bullet}{\pi}{}^{\phantom{a}\rho}_{\mu}\,\stackrel{\bullet}{\pi}{}_{\rho}^{\phantom{a}\nu},\quad
\stackrel{\bullet}{\Omega}{}^{\phantom{a}a}_{b}=\stackrel{\bullet}{\pi}{}^{\phantom{a}a}_{c}\,\stackrel{\bullet}{\pi}^{\phantom{a}c}_{b}=
\stackrel{\bullet}{\Omega}{}^{\phantom{a}\nu}_{\mu}\,\breve{e}{}^{a}_{\phantom{a}\nu}\,\breve{e}_{b}^{\phantom{a}\mu},\\
\stackrel{\bullet}{e}_{\rho}=\stackrel{\bullet}{\pi}{}_{\rho}^{\phantom{a}\nu}\,\breve{e}_{\nu}\equiv
\stackrel{\bullet}{\partial}_{\rho}=\frac{\partial}{\partial
\stackrel{\bullet}{x}{}^{\rho}},\quad
\stackrel{\bullet}{\vartheta}{}^{\rho}=\stackrel{\bullet}{\pi}{}^{\rho}_{\phantom{a}\mu}\,\breve{\vartheta}{}^{\mu}\equiv
d\,\stackrel{\bullet}{x}{}^{\rho}.
\end{array}
\label{R52}
\end{equation}
Under a local spacetime deformation
$\stackrel{\bullet}{\pi}(\stackrel{\bullet}{x})$, the tetrad changes
according to
\begin{equation}
\begin{array}{l}
\stackrel{\bullet}{e}_{a}\,\stackrel{\bullet}{\vartheta}{}^{a}=
\stackrel{\bullet}{\Omega}{}^{a}_{\phantom{a}b}\,\breve{e}_{a}\,\breve{\vartheta}{}^{b},\quad
\stackrel{\bullet}{e}_{c}=\stackrel{\bullet}{\pi}{}^{\phantom{a}a}_{c}\,\breve{e}_{a},\quad
\stackrel{\bullet}{\vartheta}{}^{c}=\stackrel{\bullet}{\pi}{}_{\phantom{a}b}^{c}\,\breve{\vartheta}{}^{b}.
\end{array}
\label{R55}
\end{equation}
Since we are interested only in a peculiar condition (\ref{R49}) to
be held, then it is completely satisfactory for further
consideration to write the first deformation matrix,
$\stackrel{\bullet}{\pi}(\stackrel{\bullet}{x})$, in the form of a
particular solution to (\ref{R49}). To derive this solution,  we
recall that for an arbitrary matrix M (\cite{rf25}),
\begin{equation}
\begin{array}{l}
tr\left\{M^{-1}\partial_{\mu}\,M\right\}=\partial_{\mu}\,\ln\,|M|,
\end{array}
\label{RW}
\end{equation}
where $|...|$ denotes the determinant, $tr$ the trace. According to
it, in matrix notation
$\breve{\omega}_{\mu}:\,=(\breve{\omega}{}^{a}_{\phantom{a}b\mu})$,
the equation (\ref{R49}) becomes
\begin{equation}
\begin{array}{l}
tr\left\{\stackrel{\bullet}{\pi}(\stackrel{\bullet}{x})\,\stackrel{\bullet}{\partial}_{\mu}\,
\stackrel{\bullet}{\pi^{-1}}(\stackrel{\bullet}{x})\right\}=-\stackrel{\bullet}{\partial}_{\mu}\ln
|\stackrel{\bullet}{\pi}(\stackrel{\bullet}{x})|=tr\,\breve{\omega}_{\mu}(\breve{x}),
\end{array}
\label{R51}
\end{equation}
which gives
\begin{equation}
\begin{array}{l}
|\stackrel{\bullet}{\pi}(\stackrel{\bullet}{x})|=
|\stackrel{\bullet}{\pi}(0)|\,
\exp\left\{-\int^{\stackrel{\bullet}{x}}_{0}\,
tr\,\breve{\omega}_{\mu}(\breve{x})\,\,d\stackrel{\bullet}{x}{}^{\,\prime\,\mu}\right\}.
\end{array}
\label{R54W}
\end{equation}
A particular solution to (\ref{R49}) is then
\begin{equation}
\begin{array}{l}
\stackrel{\bullet}{\pi}(\stackrel{\bullet}{x})=
\stackrel{\bullet}{\pi}(0)\,
\exp\left[-\int^{\stackrel{\bullet}{x}}_{0}\,
\breve{\omega}_{\mu}(\breve{x})\,d\,\stackrel{\bullet}{x}{}^{\,\prime\,\mu}\right].
\end{array}
\label{R54}
\end{equation}
This is not generally the case. However, a general solution can be
obtained by replacing $\stackrel{\bullet}{\pi}(0)\rightarrow
\pi_{B}(\stackrel{\bullet}{x})\equiv
\,\stackrel{\bullet}{\pi}(0)\,B(\stackrel{\bullet}{x})$ in the
expression (\ref{R54}), where $B(\stackrel{\bullet}{x})$ is any
proper matrix: $|B(\stackrel{\bullet}{x})|=1$. Before we report on
the further key points of physical foundation of post-Riemannian
geometry that have been used, for the benefit of the reader, we turn
back to discussion of the geometrical implications of equation
(\ref{R54}) which resembles the exponential of a bivector. Recall
that the bivectors are quantities from geometric algebra, clifford
algebra and the exterior algebra, which are generated by the
exterior product on vectors. They are used to generate rotations in
any dimension through the exponential map, and are a useful tool for
classifying such rotations, see e.g. \cite{rf26}. All bivectors in
four dimensions can be generated using at most two exterior products
and four vectors. In the case of spacetime rotations, the geometric
algebra is ${Cl}_{3,1}(R)$, and the subspace of bivectors is $\wedge
2R_{3,1}$. Accordingly, the exponential map (\ref{R54}) generates
set of all arbitrary rotations (\ref{R55}) of the orthonormal frame
$\breve{e}_{a}(\breve{x})$ in tangent space, which form the Lorentz
group. On the other hand, the universality of gravitation allows the
Levi-Civita connection to be interpreted as part of the spacetime
definition. The form of the Riemannian connection~(\ref{R27T}),
which is a function of tetrad fields and their derivatives, shows
that the relative orientation of the orthonormal frame
$\breve{e}_{a} (\breve{x}+d\,\breve{x})$ with respect to
$\breve{e}_{a}(\breve{x})$(parallel transported to
$(\breve{x}+d\,\breve{x})$ is completely fixed by the metric. Since
a change in this orientation is described by Lorentz
transformations, it does not induce any gravitational effects;
therefore, from the point of view of the principle of equivalence,
there is no reason to prevent independent (due to arbitrary
deformations (\ref{R55})) Lorentz rotations of local frames in the
space under consideration. If we want to use this freedom, the spin
connection should contain a part which is independent of the metric,
which will realize an independent Lorentz rotation of frames under
parallel transport. In this way, we are led to a description of
gravity which is not in Riemann space. If all inertial frames at a
given point are treated on an equal footing, the spacetime has to
have torsion, which is the antisymmetric part of the affine
connection. The concept of a linear connection as an independent and
primary structure of spacetime is the fundamental proposal put
forward by \'{E}lie Cartan's geometrical analysis~\cite{rf23}.
Another remark on the form of more generic spacetime deformation,
$\pi(x)$, not subject to the condition (\ref{R49}), is also in order
to validate our peculiar choice: when torsion is nonvanishing, the
affine connection is no longer coincident with the Levi-Civita
connection, and the geometry is no longer Riemannian, but one has a
Riemann-Cartan $U_{4}$ spacetime, with a nonsymmetric, but
metric-compatible, connection. Teleparallel gravity, in turn,
represented a new way of including torsion into general relativity,
an alternative to the scheme provided by the usual
Einstein-Cartan-Sciama-Kibble approach. However, for a specific
choice of free parameters, teleparallel gravity shows up as a theory
completely equivalent to Einstein's general relativity, in which
case it is usually referred to as the teleparallel equivalent of
general relativity. From this point of view, curvature and torsion
are simply alternative ways of describing the gravitational field,
and consequently related to the same degrees of freedom of gravity.
The fundamental difference between these two theories above was
that, whereas in the former torsion is a propagating field, in the
latter it is not, a point which can be considered a drawback of this
model. Therefore, we have to separate, from the very outset, these
two different cases. This motivates our choice of a double
deformation map, with the peculiar condition (\ref{R49}). Namely, we
deal with the spacetime deformation $\pi(x)$, to be consisted of two
ingredient deformations
($\stackrel{\bullet}{\pi}(\stackrel{\bullet}{x})$, \,$\sigma(x)$).
By virtue of (\ref{R49}) or (\ref{R54}), the general deformed spin
connection vanishes:
\begin{equation}
\begin{array}{l}
\stackrel{\bullet}{\omega}{}^{a}_{\phantom{a}b\mu}=\stackrel{\bullet}{\pi}{}^{\phantom{a}a}_{c}\,
\breve{\omega}{}^{c}_{\phantom{a}d\mu}\,
\stackrel{\bullet}{\pi}{}_{\phantom{b}b}^{d}+
\stackrel{\bullet}{\pi}{}^{\phantom{a}a}_{c}\,
\stackrel{\bullet}{\partial}_{\mu}\,
\stackrel{\bullet}{\pi}{}_{\phantom{b}b}^{c}=\stackrel{\bullet}{e}{}_{\phantom{a}\sigma}^{a}\,\stackrel{\bullet}{\Gamma}{}^{\sigma}_{\phantom{a}\rho\mu}
\stackrel{\bullet}{e}{}^{\phantom{b}\rho}_{b}+
\stackrel{\bullet}{e}{}_{\phantom{a}\rho}^{a}\,
\stackrel{\bullet}{\partial}_{\mu}\,
\stackrel{\bullet}{e}{}^{\phantom{b}\rho}_{b} \equiv 0.
\end{array}
\label{R56}
\end{equation}
In fact, a general linear connection,
$\stackrel{\bullet}{\Gamma}{}^{\mu}_{\phantom{a}\rho\sigma}$, is
related to the corresponding spin connection,
$\stackrel{\bullet}{\omega}{}^{a}_{\phantom{a}b\mu}$, through the
inverse
\begin{equation}
\begin{array}{l}
\stackrel{\bullet}{\Gamma}{}^{\mu}_{\phantom{a}\rho\sigma}=
\stackrel{\bullet}{e}{}^{\phantom{a}\mu}_{a}\,
\stackrel{\bullet}{\partial}_{\sigma}\,
\stackrel{\bullet}{e}{}_{\phantom{b}\rho}^{a}+
\stackrel{\bullet}{e}{}^{\phantom{a}\mu}_{a}\,
\stackrel{\bullet}{\omega}{}^{a}_{\phantom{a}b\sigma}\,
\stackrel{\bullet}{e}{}_{\phantom{b}\rho}^{b}=
\stackrel{\bullet}{e}{}^{\phantom{a}\mu}_{a}\,
\stackrel{\bullet}{\partial}_{\sigma}\,
\stackrel{\bullet}{e}{}_{\phantom{b}\rho}^{a},
\end{array}
\label{R57}
\end{equation}
which is the the Weitzenb\"{o}ck connection revealing the
Weitzenb\"{o}ck spacetime $W_{4}$ of the teleparallel gravity (see
next the section). Thus,
$\stackrel{\bullet}{\pi}(\stackrel{\bullet}{x})$ can be referred to
as the Weitzenb\"{o}ck deformation matrix. The Weitzenb\"ock
connection is a connection presenting a non-vanishing torsion, but a
vanishing curvature. This recovers a particular case of the
teleparallel gravity theory with the dynamical torsion. All
magnitudes related with the teleparallel gravity will be denoted
with an over '$\bullet$'. Furthermore,  we will be able to
generalize the EC equations for which the spin generates a dynamical
torsion part (section 4), associated with spacetime deformation
$\sigma(x)$, in the canonical energy-momentum tensor producing a
deviation from the Riemannian geometry. Equations (\ref{R56}) and
(\ref{R57}) are simply different ways of expressing the property
that the total---that is, acting on both indices---derivative of the
tetrad vanishes identically. According to the TSSD map, the next
first deformation matrices
$\sigma(x):\phantom{a}=(\sigma^{\phantom{a}a}_{b})(x)$, contribute
to corresponding ingredient part, $\chi^{\phantom{a}d}_{b}$, of the
general deformation tensor:
\begin{equation}
\begin{array}{l}
\Omega{}^{\phantom{a}a}_{b}=\chi^{\phantom{a}d}_{b}\,\stackrel{\bullet}{\Omega}{}^{\phantom{a}a}_{d}=
\chi^{\phantom{a}d}_{b}\,
\stackrel{\bullet}{\widetilde{\Omega}}{}^{\phantom{a}\nu}_{\rho}\,\breve{e}{}^{\phantom{a}a}_{\nu}\,\breve{e}_{\phantom{a}d}^{\rho},\quad
\overline{\chi}{}^{\phantom{a}c}_{d}=
\sigma^{\phantom{a}c}_{e}\,\sigma^{\phantom{a}e}_{d},\quad
\overline{\chi}{}^{\phantom{a}d}_{e}\,
\stackrel{\bullet}{\pi}{}^{\phantom{a}e}_{b}=
\chi^{\phantom{a}e}_{b}
\stackrel{\bullet}{\pi}{}^{\phantom{a}d}_{e},
\end{array}
\label{R58}
\end{equation}
or
\begin{equation}
\begin{array}{l}
\Omega^{\phantom{a}\nu}_{\mu}=
\chi^{\phantom{a}\rho}_{\mu}\,\stackrel{\bullet}{\widetilde{\Omega}}{}^{\phantom{a}\nu}_{\rho},
\quad
\chi^{\phantom{a}\rho}_{\mu}=\chi^{\phantom{a}d}_{b}\,\breve{e}{}^{\phantom{a}\rho}_{d}\,\breve{e}{}^{b}_{\phantom{a}\mu}.
\end{array}
\label{R59}
\end{equation}
Under a deformation, $\sigma(x)$, in general, the tetrad changes
according to
\begin{equation}
\begin{array}{l}

e_{c}=(\sigma^{\phantom{a}d}_{c}\,\stackrel{\bullet}{\pi}{}^{\phantom{a}a}_{d})\,\breve{e}_{a}=
\sigma^{\phantom{a}d}_{c}\,\stackrel{\bullet}{e}_{d},\quad
\vartheta^{c}=(\sigma^{c}_{\phantom{a}e}\,\stackrel{\bullet}{\pi}{}^{e}_{\phantom{a}b})\,\breve{\vartheta}{}^{b}=
\sigma^{c}_{\phantom{a}e}\,\stackrel{\bullet}{\vartheta}{}^{e},
\quad e_{\rho}=
\sigma^{\phantom{a}\sigma}_{\rho}\,\stackrel{\bullet}{e}_{\sigma},\\
\vartheta^{\rho}=
\sigma^{\phantom{a}\rho}_{\sigma}\,\stackrel{\bullet}{\vartheta}{}^{\sigma},\quad
e_{\rho}=
\sigma^{\phantom{a}c}_{\rho}\,\stackrel{\bullet}{e}_{c},\quad
\vartheta^{\rho}=
\sigma^{\phantom{a}\rho}_{c}\,\stackrel{\bullet}{\vartheta}{}^{c},\quad
\sigma^{\phantom{a}c}_{\rho}=
\sigma^{\phantom{a}\sigma}_{\rho}\,\stackrel{\bullet}{e}{}^{\phantom{a}c}_{\sigma},\quad
\sigma^{\rho}_{\phantom{a}c}=
\sigma_{\phantom{a}\sigma}^{\rho}\,\stackrel{\bullet}{e}{}^{\sigma}_{\phantom{a}c},\\

e_{c}\vartheta^{c}=\overline{\chi}{}^{\phantom{a}c}_{d}\,\stackrel{\bullet}{e}_{c}\,\stackrel{\bullet}{\vartheta}{}^{d}=
\Omega{}^{\phantom{a}a}_{b}\,\breve{e}_{a}\,\breve{\vartheta}{}^{b}.
\end{array}
\label{R61}
\end{equation}
The corresponding second deformation matrices read
\begin{equation}
\begin{array}{l}
\gamma_{cd}(x)=
\overline{\chi}_{ee'}\,\stackrel{\bullet}{\pi}{}_{c}^{\phantom{b}e}\,\stackrel{\bullet}{\pi}{}_{d}^{\phantom{b}e'},\quad
\stackrel{\bullet}{\gamma}_{cd}(\stackrel{\bullet}{x})=o_{ab}\,\stackrel{\bullet}{\pi}{}_{c}^{\phantom{b}a}(\stackrel{\bullet}{x})\,
\stackrel{\bullet}{\pi}{}_{d}^{\phantom{b}b}(\stackrel{\bullet}{x}),
\end{array}
\label{R62}
\end{equation}
where
$\overline{\chi}_{ee'}=o_{ab}\,\sigma^{\phantom{a}a}_{e}\,\sigma_{e'}^{\phantom{a}b}.$
Under a local tetrad deformation (\ref{R61}), a general spin
connection transforms according to
\begin{equation}
\begin{array}{l}
\omega'^{a}_{\phantom{a}b\mu}=\sigma{}^{\phantom{a}a}_{c}\,
\stackrel{\bullet}{\omega}{}^{c}_{\phantom{a}d\mu}\,
\sigma{}_{\phantom{b}b}^{d}+ \sigma{}^{\phantom{a}a}_{c}\,
\partial_{\mu}\, \sigma{}_{\phantom{b}b}^{c},
\end{array}
\label{R63}
\end{equation}
such that
\begin{equation}
\begin{array}{l}
\stackrel{(\sigma)}{\omega}{}^{a}_{\phantom{a}b\mu}:\phantom{a}=\omega'^{a}_{\phantom{a}b\mu}=\sigma{}^{\phantom{a}a}_{c}\,
\partial_{\mu} \,\sigma{}_{\phantom{b}b}^{c},
\end{array}
\label{RR64}
\end{equation}
is referred to as the {\em deformation related frame connection},
which represents the {\em deformed properties of the frame} only.
Then, it follows that the affine connection, $\Gamma$, related to
~(\ref{R30}) and~(\ref{R61}) tetrad deformations, transforms through
\begin{equation}
\begin{array}{l}
\Gamma^{\mu}_{\phantom{a}\rho\sigma}= e^{\phantom{a}\mu}_{a}\,
\partial_{\sigma}\, e_{\phantom{b}\rho}^{a}+
e^{\phantom{a}\mu}_{a}\,
\stackrel{(\pi)}{\omega}{}^{a}_{\phantom{a}b\sigma}\,
e_{\phantom{b}\rho}^{b}=\sigma^{\phantom{a}\mu}_{a}\,
\partial_{\sigma}\, \sigma_{\phantom{b}\rho}^{a}+
\sigma^{\phantom{a}\mu}_{a}\,
\stackrel{(\sigma)}{\omega}{}^{a}_{\phantom{a}b\sigma}\,
\sigma_{\phantom{b}\rho}^{b},
\end{array}
\label{R64T}
\end{equation}
where, according to (\ref{R61}), we have
$\sigma^{\phantom{a}\mu}_{a}\,\sigma^{\phantom{a}b}_{\mu}=\delta^{b}_{a}$,
also the procedure can be inverted
$\sigma^{\phantom{a}\mu}_{a}\,\sigma^{\phantom{a}a}_{\nu}=\delta^{\mu}_{\nu}$,
and that
\begin{equation}
\begin{array}{l}
\stackrel{(\pi)}{\omega}{}^{a}_{\phantom{a}b\mu}:\phantom{a}=\omega^{a}_{\phantom{a}b\mu}=
\pi^{a}_{\phantom{a}c}\,\breve{\omega}{}^{c}_{\phantom{a}d\mu}\,\pi^{db}+
\pi^{a}_{\phantom{a}c}\,\partial_{\mu}\,\pi^{cb},
\end{array}
\label{RR65}
\end{equation}
is the spin connection. Taking into account (\ref{R28}), observe
that invariants such as the line element, $d\,s^{2}$, defined on the
$\mathcal{M}_{4}$ by metric (\ref{R34})  can be alternatively
written in a general form of the spacetime or frame objects,
respectively, as
\begin{equation}
\begin{array}{l}
 d\,s^{2}=g_{\mu\nu}\,\vartheta^{\mu}\otimes
\vartheta^{\nu}=g(e_{\mu},e_{\nu})\,\vartheta^{\mu}\otimes
\vartheta^{\nu}=
\left(\Omega_{\mu}^{\phantom{a}\nu}\,\Omega_{\rho}^{\phantom{a}\sigma}\right)\,\breve{g}_{\nu\sigma}
\,\breve{\vartheta}{}^{\mu}\otimes\breve{\vartheta}{}^{\rho}=
o_{ab}\,\vartheta^{a}\otimes \vartheta^{b}=\\
\left(\Omega_{a}^{\phantom{a}c}\,\Omega_{b}^{\phantom{a}d}\right)
o_{cd}\,\breve{\vartheta}{}^{a}\otimes \,\breve{\vartheta}{}^{b}=
\gamma_{cd}\,\breve{\vartheta}{}^{c}\otimes
\,\breve{\vartheta}{}^{d}.
\end{array}
\label{R65}
\end{equation}
For our convenience, hereinafter the notation,
$\{\stackrel{(A)}{e}_{a},\,\stackrel{(A)}{\vartheta}{}^{b}\}\,
(A=\pi, \sigma)$, will be used for general linear frames
\begin{equation}
\begin{array}{l}
\{\stackrel{(A)}{e}_{a},\,\stackrel{(A)}{\vartheta}{}^{b}\}=
\{(\stackrel{(\pi)}{e}_{a},\,\stackrel{(\sigma)}{e}_{a}),\,
(\stackrel{(\pi)}{\vartheta}{}^{b},\,\stackrel{(\sigma)}{\vartheta}{}^{b})\}\equiv
\{(e_{a},\,\stackrel{\bullet}{e}_{a}),\,
(\vartheta^{b},\,\stackrel{\bullet}{\vartheta}{}^{b})\},
\end{array}
\label{F1}
\end{equation}
where  $\stackrel{(A)}{e}_{a}\,\rfloor\,
\stackrel{(A)}{\vartheta}{}^{b}=\delta^{b}_{a}$, or in components,
$\stackrel{(A)}{e}{}_{a}^{\phantom{a}\mu}\,
\stackrel{(A)}{e}{}^{b}_{\phantom{a}\mu}=\delta^{b}_{a}$, also the
procedure can be inverted
$\stackrel{(A)}{e}{}^{a}_{\phantom{a}\rho}\,\stackrel{(A)}{e}{}_{a}^{\phantom{a}
\sigma}=\delta^{\sigma}_{\rho}$,  provided
\begin{equation}
\begin{array}{l}
\stackrel{(A)}{e}{}_{a}^{\phantom{a}\mu}=
(\stackrel{(\pi)}{e}{}_{a}^{\phantom{a}\mu},\,\stackrel{(\sigma)}{e}{}_{a}^{\phantom{a}\mu})\equiv
(e_{a}^{\phantom{a}\mu},\,\sigma_{a}^{\phantom{a}\mu}).
\end{array}
\label{F2}
\end{equation}
Hence, the affine connection (\ref{R64T}) can be rewritten in the
abbreviated form
\begin{equation}
\begin{array}{l}
\Gamma^{\mu}_{\phantom{a}\rho\sigma}=
\stackrel{(A)}{e}{}^{\phantom{a}\mu}_{a}\,
\partial_{\sigma}\, \stackrel{(A)}{e}{}_{\phantom{b}\rho}^{a}+
\stackrel{(A)}{e}{}^{\phantom{a}\mu}_{a}\,
\stackrel{(A)}{\omega}{}^{a}_{\phantom{a}b\sigma}\,
\stackrel{(A)}{e}{}_{\phantom{b}\rho}^{b}.
\end{array}
\label{F3}
\end{equation}
Since the first deformation matrices $\pi(x)$ and $\sigma(x)$ are
arbitrary functions, the transformed general spin connections
$\stackrel{(\pi)}\omega(x)$ and $\stackrel{(\sigma)}\omega(x)$, as
well as the affine connection~(\ref{F3}), are independent of tetrad
fields and their derivatives. In what follows, therefore, we will
separate the notions of space and connections- the metric-affine
formulation of gravity. A  metric-affine space $(\mathcal{M}_{4},$
$\,g,\,\Gamma)$ is defined to have a metric and a linear connection
that need not dependent on each other. The new geometrical property
of the spacetime are the {\em nonmetricity} 1-form $N_{ab}$ and the
affine {\em torsion} 2-form $T^{a}$ representing a translational
misfit (for a comprehensive discussion see~\cite{rf9,rf10,rf13}.
These, together with the {\em curvature} 2-form
$R_{a}^{\phantom{a}b}$, symbolically can be presented as \cite{rf6}
\begin{equation}
\begin{array}{l}
\left(N_{ab},\,T^{a},\,R_{a}^{\phantom{a}b}\right)\;\sim\; {\cal
D}\left(g_{ab},\,\vartheta^{a},\, \Gamma_{a}^{\phantom{a}b}\right),
\end{array}
\label{R66}
\end{equation}
where ${\cal D}$ is the {\it covariant exterior derivative}. If the
nonmetricity tensor $N_{\lambda\mu\nu}=-{\cal
D}_{\lambda}\,g_{\mu\nu}\equiv -g_{\mu\nu\,; \lambda}$ does not
vanish, the general formula for the affine connection written in the
spacetime components is~\cite{rf13}
\begin{equation}
\begin{array}{l}
\Gamma^{\rho}_{\phantom{a}\mu\,\nu}=\stackrel{\circ
}{\Gamma}{}^{\rho}_{\phantom{a}\mu\,\nu}
+K^{\rho}_{\phantom{k}\mu\nu}-N^{\rho}_{\phantom{k}\mu\nu}+\frac{1}{2}N_{(\mu\phantom{k}\nu)}^{\phantom{(i}\rho},
\end{array}
\label{RU1}
\end{equation}
where
$K^{\rho}_{\phantom{k}\mu\nu}:\phantom{a}=2Q_{(\mu\nu)}^{\phantom{(ij)}\rho}+
Q^{\rho}_{\phantom{k}\mu\nu}$ is the non-Riemann part - the affine
{\em contortion tensor}.  The torsion,
$Q^{\rho}_{\phantom{k}\mu\nu}=
\frac{1}{2}\,T^{\rho}_{\phantom{k}\mu\nu}=\Gamma^{\rho}_{\phantom{a}[\mu\,\nu]}$
given with respect to a holonomic frame, $d\,\vartheta^{\rho}=0$, is
a third-rank tensor, antisymmetric in the first two indices, with 24
independent components. In the presence of curvature and torsion,
the coupling prescription of a general field carrying an arbitrary
representation of the Lorentz group will be
\begin{equation}
\begin{array}{l}
\partial_{\mu} \rightarrow {\cal D}_{\mu}=
\partial_{\mu} - \frac{i}{2}
\left(\omega^{ab}_{\phantom{a}\mu} - K^{ab}_{\phantom{a}\mu} \right)
J_{ab},
\end{array}
\label{R70A}
\end{equation}
We may introduce the contortion tensors related to the {\em
deformation-related frame connection} (\ref{RR64}) and the spin
connection (\ref{RR65}):
\begin{equation}
\begin{array}{l}
\stackrel{(A)}{K}{}^{c}_{\phantom{a}a\nu}=\stackrel{(A)}{\omega}{}^{c}_{\phantom{a}a\nu}+
\stackrel{(A)}{\Delta}{}^{c}_{\phantom{a}a\nu},,
\end{array}
\label{R73}
\end{equation}
where
\begin{equation}
\begin{array}{l}
\stackrel{(A)}\Delta_{\mu\rho\nu}=\stackrel{(A)}{e}_{\mu
a}\,\stackrel{(A)}{e}{}^{a}_{[\rho,\,\nu]}-\stackrel{(A)}{e}_{\rho
a}\,\stackrel{(A)}{e}{}^{a}_{[\mu,\,\nu]}-\stackrel{(A)}{e}_{\nu
a}\,\stackrel{(A)}{e}{}^{a}_{[\mu,\,\rho]},
\end{array}
\label{E8}
\end{equation}
is referred to as the the Ricci coefficients of rotation. Both the
contortion tensor and spin connection are antisymmetric in their
first two indices. The Levi-Civita spin connection
\begin{equation}
\begin{array}{l}
\stackrel{\circ \,(A)
}{\omega}{}^{\mu}_{\phantom{i}a\rho}=\stackrel{(A)}{e}{}^{\mu}_{a\,:\rho}=\stackrel{(A)}{e}{}^{\mu}_{a,\,\rho}+
\stackrel{\circ}{\Gamma}{}^{\mu}_{\phantom{a}\nu\rho}\,\stackrel{(A)}{e}{}^{\nu}_{a},
\end{array}
\label{E9}
\end{equation}
is related to the Ricci rotation coefficients, with
$\stackrel{(A)}{K}=0$, thus,
\begin{equation}
\begin{array}{l}
\stackrel{(A)}{K}{}^{\mu}_{\phantom{i}a\rho}=\stackrel{(A)}{\omega}{}^{\mu}_{\phantom{i}a\rho}-
\stackrel{\circ\,(A)}{\omega}{}^{\mu}_{\phantom{i}a\rho}.
\end{array}
\label{E10}
\end{equation}
The relations between the corresponding torsion and contortion
tensors read
\begin{equation}
\begin{array}{l}
\stackrel{(A)}{K}{}^{\rho}_{\phantom{k}\mu\nu}:\phantom{a}=2\stackrel{(A)}{Q}{}_{(\mu\nu)}^{\phantom{(ij)}\rho}+
\stackrel{(A)}{Q}{}^{\rho}_{\phantom{k}\mu\nu}, \quad
\stackrel{(A)}{Q}{}^{\rho}_{\phantom{k}\mu\nu}=\stackrel{(A)}{K}{}^{\rho}_{\phantom{k}[\mu\nu]},
\end{array}
\label{E11}
\end{equation}
where
\begin{equation}
\begin{array}{l}
\stackrel{(A)}{Q}{}^{\rho}_{\phantom{k}\mu\nu}=\stackrel{(A)}{\omega}{}^{\rho}_{\phantom{k}[\mu\nu]}+
\stackrel{(A)}{e}{}^{a}_{\phantom{k}[\mu,\,\nu]}\,\stackrel{(A)}{e}{}_{a}^{\rho}.
\end{array}
\label{E12}
\end{equation}
So, the affine connection (\ref{F3}) reads
\begin{equation}
\begin{array}{l}
\Gamma^{\mu}_{\phantom{a}\rho\sigma}=\stackrel{(\sigma)}{\Gamma}{}^{\mu}_{\phantom{a}(\rho\sigma)}+\stackrel{(\sigma)}{Q}{}^{\mu}_{\phantom{a}\rho\sigma}=
\stackrel{(\pi)}{\Gamma}{}^{\mu}_{\phantom{a}(\rho\sigma)}+\stackrel{(\pi)}{Q}{}^{\mu}_{\phantom{a}\rho\sigma},
\end{array}
\label{E13}
\end{equation}
where
$\stackrel{(A)}{\Gamma}{}^{\mu}_{\phantom{a}(\rho\sigma)}=\stackrel{\circ}{\Gamma}{}^{\mu}_{\phantom{a}(\rho\sigma)}+
2\stackrel{(A)}{Q}{}^{\mu}_{\phantom{a}(\rho\sigma)}$. It is well
known that due to the affine character of the connection space
\cite{rf22}, one can always add a tensor to a given connection
without spoiling the covariance of derivative~(\ref{R70A}). Let us
define then a translation in the connection space. Suppose a point
in this space will be a Lorentz connection, $
\stackrel{(\pi)}{\omega}(x):\phantom{a}=\stackrel{(\pi)}{\omega}{}^{bc}_{\phantom{aa}\mu}(x)\,J_{bc}\,d\,x{}^{\mu},
$ presenting simultaneously curvature and torsion written in the
language of differential forms as
\begin{equation}
\begin{array}{l}
\stackrel{(\pi)}{R} = d\, \stackrel{(\pi)}{\omega} +
\stackrel{(\pi)}{\omega} \, \stackrel{(\pi)}{\omega} \equiv {\cal
D}_{\stackrel{(\pi)}{\omega}} \,\stackrel{(\pi)}{\omega},\quad
\stackrel{(\pi)}{T} = d\, e +  \stackrel{(\pi)}{\omega} \,e \equiv
{\cal D}_{\stackrel{(\pi)}{\omega}} \, e,
\end{array}
\label{RU14}
\end{equation}
where  ${\cal D}_{\stackrel{(\pi)}{\omega}}$ denotes the covariant
differential in the connection $\stackrel{(\pi)}{\omega}$. Now,
given two connections $\stackrel{(\sigma)}{\omega}(x)$ and
$\stackrel{(\pi)}{\omega}(x)$, the difference $
k=\stackrel{(\sigma)}{\omega}-\stackrel{(\pi)}{\omega}$, is also a
1-form assuming values in the Lorentz Lie algebra, but transforming
covariantly, whereas its covariant derivative is $ {\cal
D}_{\stackrel{(\pi)}{\omega}} \,k=d\, k+ \{\stackrel{(\pi)}{\omega},
\,k\}. $ The effect of adding a covector $k$ to a given connection
$\stackrel{(\pi)}\omega$, therefore, is to change its curvature and
torsion 2-forms:
\begin{equation}
\begin{array}{l}
\stackrel{(\sigma)}{R}=\stackrel{(\pi)}{R}+{\cal
D}_{\stackrel{(\pi)}{\omega}} \,k +k\,k,\quad
\stackrel{(\sigma)}{T}=\stackrel{(\pi)}{T}+k\,e.
\end{array}
\label{RU17}
\end{equation}
Since $k_{\phantom{a}bc}^{a}$ is a Lorentz-valued covector, it is
necessarily anti-symmetric in the first two indices. Presenting $
k_{\phantom{a}bc}^{a}=\frac{1}{2}\,k_{\phantom{a}(bc)}^{a}+\frac{1}{2}\,k_{\phantom{a}[bc]}^{a},
$ we may define $k_{\phantom{a}[bc]}^{a}\equiv
t_{\phantom{a}bc}^{a}, $ such that
\begin{equation}
\begin{array}{l}
k^a{}_{bc} = \frac{1}{2} (t_a{}^c{}_b + t_b{}^c{}_a - t^a{}_{bc} ).
\end{array}
\label{RU18}
\end{equation}
Turning to the connection appearing in the covariant
derivative~(\ref{R70A}): $ \stackrel{(\pi)}{\Omega}{}^a{}_{bc}
\equiv \stackrel{(\pi)}{\omega}{}^a{}_{bc} -
\stackrel{(\pi)}{K}{}^a{}_{bc}, $ a translation in the connection
space with parameter $k^a{}_{bc}$ corresponds to
\begin{equation}
\begin{array}{l}
\stackrel{(\sigma)}{\Omega}{}^a_{\phantom{a}bc} =
\stackrel{(\pi)}{\Omega}{}^a_{\phantom{a}bc} + k^a_{\phantom{a}bc}
\equiv \stackrel{(\pi)}{\omega}{}^a_{\phantom{a}bc} -
\stackrel{(\pi)}{K}{}^a{}_{bc} + k^a_{\phantom{a}bc}.
\end{array}
\label{RU20}
\end{equation}
Since $k^a_{\phantom{a}bc}$ has always the form of a contortion
tensor~(\ref{RU18}), the above connection is equivalent to $
\stackrel{(\sigma)}{\Omega}{}^a_{\phantom{a}bc} =
\stackrel{(\pi)}{\omega}{}^a_{\phantom{a}bc} -
\stackrel{(\sigma)}{K}{}^a_{\phantom{a}bc}, $ with
$\stackrel{(\sigma)}{K}{}^a_{\phantom{a}bc} =
\stackrel{(\pi)}{K}{}^a_{\phantom{a}bc} - k^a_{\phantom{a}bc}$ being
another contortion tensor: $
\stackrel{(\sigma)}{K}{}^{\lambda}_{\phantom{a}\mu\nu}=\stackrel{(\pi)}{K}{}^{\lambda}_{\phantom{a}\mu\nu}-
k^{\lambda}_{\phantom{a}\mu\nu}. $ Let us, for example, in
(\ref{RU17}) choose $t^a_{\phantom{a}bc}$ as the torsion of the
connection $\stackrel{(\pi)}{\omega}{}^a_{\phantom{a}bc}$, that is,
$t^a_{\phantom{a}bc} = \stackrel{(\pi)}{T}{}^a_{\phantom{a}bc}$. In
this case, $k^a_{\phantom{a}bc} =
\stackrel{(\pi)}{K}{}^a_{\phantom{a}bc}$, and  that we are left with
the torsionless spin connection of general relativity: $
\stackrel{(\sigma)}{\Omega}{}^a_{\phantom{a}bc} =
\stackrel{\circ}{\omega}{}^a_{\phantom{a}bc}. $ Another example is $
t^a_{\phantom{a}bc} = \stackrel{(\pi)}{T}{}^a_{\phantom{a}bc} -
\stackrel{(\pi)}{C}{}^a_{\phantom{a}bc}, $ when the connection
$\stackrel{(\pi)}{\omega}{}^a_{\phantom{a}bc}$ vanishes, which
characterizes teleparallel gravity. In this case, the resulting
connection has the form $
\stackrel{(\sigma)}{\omega}{}^a_{\phantom{a}bc} = -
\stackrel{\bullet}{K}{}^a_{\phantom{a}bc}, $ where
$\stackrel{\bullet}{K}{}^a_{\phantom{a}bc}$ is the contortion tensor
written in terms of the Weitzenb\"ock torsion
$\stackrel{\bullet}{T}{}^a_{\phantom{a}bc} = -
\stackrel{\bullet}{C}{}^a_{\phantom{a}bc}$. The particle equation of
motion then becomes the force equation of teleparallel gravity.
There are actually infinitely many choices for $t^a{}_{bc}$, each
one defining a different translation in the connection space, and
consequently yielding a connection
$\stackrel{(\pi)}{\omega}{}^a_{\phantom{a}bc}$ with different
curvature and torsion.

\section{Teleparallel gravity}
The total covariant derivative of a geometrical object carrying both
flat and curvilinear indices is covariant with respect to both
diffeomorphism and local Lorentz symmetries. In both, $W_{4}$ and
$U_{4}$, spaces the total covariant derivative of the vierbein
field, $e^{a}_{\phantom{a}\nu} $, is assumed to vanish
\begin{equation}
\begin{array}{l}
{\cal D}_{\mu}\,e^{a}_{\phantom{a}\nu} =\partial_{\mu}
\,e^{a}_{\phantom{a}\nu} - \Gamma^{\rho}_{\phantom{a}\nu \mu}
\,e^{a}_{\phantom{a}\rho} + \omega^{a}_{\phantom{a}b\mu}
\,e^{b}_{\phantom{a}\nu}=0,
\end{array}
\label{RS1}
\end{equation}
which provides a relation between both connections. Defining the
Weitzenb\"{o}ck connection $ \stackrel{\bullet}{\Gamma}{}
^{\rho}_{\phantom{a}\nu \mu }=e^{\phantom{a}\rho}_{a}
\partial_{\mu}\,e^{a}_{\phantom{a}\nu}$ ((\ref{R57})), then
(\ref{RS1})
 leads to
 \begin{equation}
\begin{array}{l}
\stackrel{\bullet}{\Gamma}{} ^{\rho}_{\phantom{a}\mu \nu }= \Gamma
^{\rho}_{\phantom{a}\mu\nu}- \omega_{\phantom{a}\mu
b}^{a}\,\,e^{\phantom{a}\rho}_{a}\,e^{b}_{\phantom{a}\nu}=
\Gamma^{\rho}_{\phantom{a}\mu\nu}-\omega_{\phantom{a}\mu\nu}^{\rho}=
\stackrel{\circ}{\Gamma}{} ^{\rho}_{\phantom{a}\mu \nu
}+K^{\rho}_{\phantom{a}\mu \nu }-\omega_{\phantom{a}\mu\nu}^{\rho}.
\end{array}
\label{RC2}
\end{equation}
The anholonomy object $C^{a}$ and the torsion 2-form $T^{a}$,
expanded in the anholonomic coframe $\{\vartheta^{a}\}$, read
\begin{equation}
\begin{array}{l}
C^{a}=\frac{1}{2}C^{a}_{\phantom{a}bc}\,\vartheta^{b}\wedge
\vartheta^{c},\quad
T^{a}=\frac{1}{2}T^{a}_{\phantom{a}bc}\,\vartheta^{b}\wedge
\vartheta^{c},
\end{array}
\label{RH2}
\end{equation}
where
$T^{a}_{\phantom{a}bc}=C^{a}_{\phantom{a}bc}+\Gamma^{a}_{\phantom{a}bc}-\Gamma^{a}_{\phantom{a}cb}.
$ Let us now introduce the Weitzenb\"{o}ck torsion
$\stackrel{\bullet}{T}{} ^{\rho}_{\phantom{a}\mu \nu
}:\phantom{a}=\stackrel{\bullet}{\Gamma}{} ^{\rho}_{\phantom{a}\mu
\nu }-\stackrel{\bullet}{\Gamma}{} ^{\rho}_{\phantom{a}\nu \mu }, $
and the Weitzenb\"{o}ck contortion
\begin{equation}
\begin{array}{l}
\stackrel{\bullet}{K}{}^\rho{}_{\mu\nu}:\phantom{a} = {\textstyle
\frac{1}{2}} (\stackrel{\bullet}{T}{}_\nu{}^\rho{}_\mu +
\stackrel{\bullet}{T}{}_\mu{}^\rho{}_\nu +
\stackrel{\bullet}{T}{}^\rho{}_{\mu\nu}).
\end{array}
\label{RC6}
\end{equation}
From (\ref{RS1}), we then obtain $ \stackrel{\bullet}{T}{}
^{\rho}_{\phantom{a}\mu \nu }=T^{\rho}_{\phantom{a}\mu \nu
}-\omega_{\phantom{a}\mu\nu}^{\rho}+\omega_{\phantom{a}\nu\mu}^{\rho}.
$ and $ \stackrel{\bullet}{K}{}^\rho{}_{\phantom{a}\mu\nu}=
K^\rho{}_{\phantom{a}\mu\nu} +\omega^\rho{}_{\mu\nu}. $ Below, we
will concentrate on the specific space, $W_{4}$, of the vanishing
affine torsion in the class of frames,
$\{\stackrel{\bullet}{e}_{a}\}$:\phantom{a}
 $T^{\rho}_{\phantom{a}\mu \nu
}=0. $  Provided, the metricity condition holds: $
\stackrel{\bullet}{N}_{ab}: \phantom{a} =-\stackrel{\bullet}{\cal
D}g_{ab}=0 $, and that $ \Gamma^{\rho}_{\phantom{a}\mu
\nu}=\stackrel{\circ}{\Gamma}{} ^{\rho}_{\phantom{a}\mu \nu}, $ as
in the Riemann space. Consequently, $
K^{\rho}_{\phantom{a}\mu\nu}=0, $ and $
\stackrel{\bullet}{K}{}^\rho{}_{\mu\nu}=\omega_{\phantom{a}\mu\nu}^{\rho}.
$ Hence, the  (\ref{RS1}) yields $
\stackrel{\bullet}{\Gamma}{}^\rho{}_{\mu\nu}=\stackrel{\circ}{\Gamma}{}^\rho{}_{\mu\nu}+
\stackrel{\bullet}{K}{}^\rho{}_{\mu\nu}= \Gamma^\rho{}_{\mu\nu}+
\stackrel{\bullet}{K}{}^\rho{}_{\mu\nu}, $ while, the Weitzenb\"ock
covariant derivative of the tetrad field vanishes identically:
$\stackrel{\bullet}{\cal
D}_{\nu}\,\stackrel{\bullet}{e}{}^{a}_{\phantom{a}\mu} \equiv
\stackrel{\bullet}{\partial}_{\nu}\,\stackrel{\bullet}{e}{}^{a}_{\phantom{a}\mu}
- \stackrel{\bullet}{\Gamma}{}^{\rho}_{\phantom{a}\mu \nu} \,
\stackrel{\bullet}{e}{}^{a}_{\phantom{a}\rho} = 0. $ This is the so
called distant, or absolute parallelism condition. As a consequence
of this condition, the corresponding Weitzenb\"ock spin connection
also vanishes identically: $
\stackrel{\bullet}{\omega}{}^{c}_{\phantom{a}a\nu} =
\stackrel{\circ}{\omega}{}^{c}_{\phantom{a}a\nu} +
\stackrel{\bullet}{K}{}^{c}_{\phantom{a}a\nu}\equiv 0. $ Of course,
these relations above are true only in one specific class of frames.
In fact, since $\stackrel{\bullet}{\omega}{}^{c}_{\phantom{a}a\nu} $
is the Weitzenb\"{o}ck spin connection, if it vanishes in a given
frame, it will be different from zero in any other frame related to
the first by a local Lorentz transformation. In teleparallel
gravity, the coupling of spinor fields with gravitation is a highly
controversial subject. However, it seems there are no compelling
arguments supporting the choice of the Weitzenb\"{o}ck spin
connection $\stackrel{\bullet}{\omega}{}^{c}_{\phantom{a}a\nu}$ as
the spin connection of teleparallel gravity, otherwise  several
problems are immediately encountered  with such coupling
prescription. The teleparallel gravity becomes consistent and fully
equivalent with GR, even in the presence of spinor fields if we
write the minimal coupling prescription as $
\stackrel{\bullet}{\partial}_a \rightarrow \stackrel{\bullet}{\cal
D}_{a}= \stackrel{\bullet}{e}{}_{a}^{\phantom{a}\mu} \,
\stackrel{\bullet}{\cal D}_{\mu} $ with $\stackrel{\bullet}{\cal
D}_{\mu}$ the teleparallel Fock-Ivanenko derivative written in the
form $ \stackrel{\bullet}{\cal D}_{\mu}:\phantom{a} =
\stackrel{\bullet}{\partial}_{\mu} - \frac{i}{2} \,
\stackrel{\bullet}{\Omega}{}^{a}_{\phantom{a}b \mu}
\,J_{a}^{\phantom{a}b}, $ where the teleparallel spin connection,
$\stackrel{\bullet}{\Omega}{}^{a}_{\phantom{a}b \mu}$, reads
$\stackrel{\bullet}{\Omega}{}^{a}_{\phantom{a}b \mu}:\phantom{a}= 0
- \stackrel{\bullet}{K}{}^{a}_{\phantom{a}b \mu}. $  Field equations
can be derived from the least action, $\delta
\stackrel{\bullet}{S}=0$, with the total invariant action of
conventional theory of teleparallel gravity.

\section{In search of TSSD-induced dynamical torsion}
In this section we construct the TSSD-$U_{4}$ theory, which
considers curvature and torsion as representing independent degrees
of freedom. The RC manifold, $U_{4}$, is a particular case of the
general metric-affine manifold $\mathcal{M}_{4}$, restricted by the
metricity condition $N_{\lambda\mu\nu}=0$, when a nonsymmetric
linear connection is said to be metric compatible. Taking the
antisymmetrized derivative of the metric condition gives an identity
between the curvature of the spin-connection and the curvature of
the Christoffel connection
\begin{equation}
\begin{array}{l}
\stackrel{(A)}{R}{}_{\mu\nu}^{\phantom{ab} ab}(
\stackrel{(A)}{\omega}) \stackrel{(A)}{e}_{\rho
b}-R_{\phantom{a}\rho\mu\nu }^{\sigma}(  \Gamma)\,
\stackrel{(A)}{e}{}_{\sigma}^{\phantom{a}a}=0,
\end{array}
\label{RCC1}
\end{equation}
where
\begin{equation}
\begin{array}{l}
\stackrel{(A)}{R}{}_{\mu\nu}^{\quad ab}( \stackrel{(A)}{\omega})
=\partial_{\mu}\stackrel{(A)}{\omega}{}_{\nu
}^{\phantom{a}ab}-\partial_{\nu}\stackrel{(A)}{\omega}{}_{\mu}^{\phantom{a}ab}+\stackrel{(A)}{\omega}{}_{\mu}^{\phantom{a}ac}
\stackrel{(A)}{\omega}{}_{\nu
c}^{\phantom{a}b}-\stackrel{(A)}{\omega}{}_{\nu}^{\phantom{a}ac}\stackrel{(A)}{\omega}{}_{\mu c}^{\phantom{a}b},\\
R_{\phantom{a}\rho\mu\nu}^{\sigma}( \Gamma)
=\partial_{\mu}\Gamma_{\nu\rho}^{\sigma}-\partial_{\nu}\Gamma_{\mu\rho}^{\sigma}-
\Gamma_{\mu\rho
}^{\lambda}\Gamma_{\nu\lambda}^{\sigma}+\Gamma_{\nu\rho}^{\lambda}
\Gamma_{\mu\lambda}^{\sigma}.
\end{array}
\label{R56C}
\end{equation}
Hence, the relations between the scalar curvatures for an $U_{4}$
manifold read
\begin{equation}
\begin{array}{l}
\stackrel{(A)}{R}( \stackrel{(A)}{\omega})\equiv
\stackrel{(A)}{e}{}_{a}^{\phantom{a}\mu}\,\stackrel{(A)}{e}{}_{b}^{\phantom{a}\nu}\,\stackrel{(A)}{R}{}_{\mu\nu}^{\phantom{ab}
ab}( \stackrel{(A)}{\omega})=R(g,\,\Gamma)\equiv g^{\rho\nu
}\,R_{\;\rho\mu\nu}^{\mu}(\Gamma).
\end{array}
\label{R57C}
\end{equation}
This means that the Lorentz and diffeomorphism invariant scalar
curvature, $R$, becomes either a function of
$\stackrel{(A)}{e}{}_{\mu}^{a}$  only, or  $g_{\mu\nu}$. Certainly,
it can be seen by noting that the Lorentz gauge transformations can
be used to fix the six antisymmetric components of
$\stackrel{(A)}{e}{}_{\mu}^{a}$ to vanish. Then in both cases
diffeomorphism invariance fixes four more components out of the six
$g_{\mu\nu}$, with the four components $g_{0\mu}$ being non
dynamical, obviously, leaving only two dynamical degrees of freedom.
This shows that the equivalence of the vierbein and metric
formulations holds.

\subsection{Field equations of dynamical torsion in the tensorial form}
According to relations (\ref{R57C}), the  total {\em EC action} can
be written in the terms of the spin connection,
$\stackrel{(\pi)}{\omega}$ and the {\em deformation-related frame
connection}, $\stackrel{(\sigma)}{\omega}$, in the form
\begin{equation}
\begin{array}{l}
 S=S_{g}^{(A)}( \stackrel{(A)}{\omega}) +S_{m}^{(\pi)}(
\stackrel{(\pi)}{\omega})=-\frac{1}{2{\ae}}\,\int
\stackrel{(A)}{R}\sqrt{-g}\,d\,\Omega +\int
\,L_{m}^{(\pi)}(g,\,\Psi,\, \nabla\,\Psi)\sqrt{-g}\,d\,\Omega,
\end{array}
\label{RU24}
\end{equation}
where $S_{g}^{(A)}\quad (A=\pi,\, \sigma)$ is the action for the
gravitational field written, according to (\ref{R57C}), in terms of
scalar curvature $\stackrel{(A)}{R}( \stackrel{(A)}{\omega})$ for a
$U_{4}$ manifold, while $S_{m}^{(\pi)}$ is the action for the matter
fields, $\ae$ is the coupling constant relating to the Newton
gravitational constant $\ae=8\pi G/c^{4}$. Action (\ref{RU24})
regards the contortion tensor as a variational variable, in addition
to the gravitational and matter fields. The gravitational action can
be decomposed as
\begin{equation}
\begin{array}{l}
S_{g}^{(A)}=-\frac{1}{2{\ae}}\,\int
\stackrel{\circ}{R}\sqrt{-g}\,d\,\Omega+S_{Q}^{(A)},
\end{array}
\label{PPP2}
\end{equation}
where the torsional action  reads
\begin{equation}
\begin{array}{l}
 S_{Q}^{(A)}=\frac{1}{2{\ae}}\int
d\,\Omega\,\sqrt{-g}L_{Q}^{(A)}=\frac{1}{2{\ae}}\int
d\,\Omega\,\sqrt{-g}
g^{\mu\rho}(2\stackrel{(A)}{K}{}^{\lambda}_{\phantom{l}\mu\lambda:\rho}+
\stackrel{(A)}{K}{}^{\nu}_{\phantom{j}\mu\nu}\stackrel{(A)}{K}{}^{\lambda}_{\phantom{l}\rho\lambda}-
\stackrel{(A)}{K}{}^{\lambda}_{\phantom{l}\mu\sigma}\stackrel{(A)}{K}{}^{\sigma}_{\phantom{m}\rho\lambda}),
\end{array}
\label{RU26}
\end{equation}
The coupling constant of the spin-torsion is the same of that of the
mass-metric distortion field interaction. Partial integration of the
terms with covariant derivatives $(:)$ and omitting total
derivatives (which do not contribute to the field equations) reduces
the action $S_{Q}^{(A)}$  to
\begin{equation}
\begin{array}{l}
S_{Q}^{(A)}=\frac{1}{2{\ae}}\,\int d\,\Omega\,\sqrt{-g}\,
g^{\mu\rho}\,(
\stackrel{(A)}{K}{}^{\nu}_{\phantom{j}\mu\nu}\,\stackrel{(A)}{K}{}^{\lambda}_{\phantom{l}\rho\lambda}-
\stackrel{(A)}{K}{}^{\lambda}_{\phantom{l}\mu\sigma}\,\stackrel{(A)}{K}{}^{\sigma}_{\phantom{m}\rho\lambda}).
\end{array}
\label{RP26}
\end{equation}
The corresponding variations can be written as
\begin{equation}
\begin{array}{l}
\delta\,S_{Q}^{(A)}=\frac{1}{2{\ae}}\,\int[\stackrel{(A)}{K}{}^{\nu}_{\phantom{j}\mu\nu}\,
\stackrel{(A)}{K}{}^{\lambda}_{\phantom{l}\rho\lambda}-
\stackrel{(A)}{K}{}^{\lambda}_{\phantom{l}\mu\sigma}\,
\stackrel{(A)}{K}{}^{\sigma}_{\phantom{m}\rho\lambda}-
\frac{1}{2}\,g_{\mu\rho}\,(\stackrel{(A)}{K}{}^{\nu\sigma}_{\phantom{jm}\nu}\,
\stackrel{(A)}{K}{}^{\lambda}_{\phantom{l}\sigma\lambda}-\\
\stackrel{(A)}{K}{}^{\lambda\nu}_{\phantom{lj}\sigma}\,\stackrel{(A)}{K}{}^{\sigma}_{\phantom{m}\nu\lambda})]
\,\sqrt{-g}\,\delta
g^{\mu\rho}\,d\,\Omega\,-\frac{1}{{\ae}}\int(\stackrel{(A)}{K}{}^{\rho\nu}_{\phantom{kj}\mu}-
\stackrel{(A)}{K}{}^{\lambda\nu}_{\phantom{lj}\lambda}\,\delta^{\rho}_{\mu})\,\sqrt{-g}\,\delta
\stackrel{(A)}{K}{}^{\mu}_{\phantom{i}\nu\rho}\,d\,\Omega,
\end{array}
\label{E1}
\end{equation}
and
\begin{equation}
\begin{array}{l}
\delta\,S_{m}^{(\pi)}=\frac{1}{2}\int T_{\mu\rho}\,\sqrt{-g}\,\delta
g^{\mu\rho}\,d\,\Omega +\frac{1}{2}\int
\stackrel{(\pi)}{S}{}^{\phantom{a}\mu\rho}_{\nu}\,\sqrt{-g}\,\delta
\stackrel{(\pi)}{\omega}{}^{\nu}_{\phantom{j}\mu\rho}\,d\,\Omega,
\end{array}
\label{E2}
\end{equation}
where $T_{\mu\rho}$ is the usual dynamical energy-momentum tensor,
and $\stackrel{(\pi)}{S}{}^{\phantom{a}\mu\rho}_{\nu}$ is the spin
tensor. In the metric-affine variational formulation of gravity, the
variations $\delta
\stackrel{(\pi)}{\omega}{}^{ab}_{\phantom{ab}\mu}$ are independent
of $\delta e^{\phantom{a}\mu}_{a}$ and their derivatives. The
dynamical spin density tensor, which is antisymmetric in the Lorentz
indices, reads
\begin{equation}
\begin{array}{l}
\stackrel{(\pi)}{s}{}^{\phantom{a}ab}_{\mu}=2\,\frac{\delta
(\sqrt{-g}\,L_{m}^{(\pi)})}{\delta
\stackrel{(\pi)}{\omega}{}^{\mu}_{\phantom{j}ab}}= 2\,\frac{\delta
(\sqrt{-g}\,L_{m}^{(\pi)})}{\delta
\stackrel{(\pi)}{K}{}^{\mu}_{\phantom{j}ab}}=
\sqrt{-g}\,\stackrel{(\pi)}{S}{}^{\phantom{a}ab}_{\mu},
\end{array}
\label{RU31}
\end{equation}
where (\ref{R73}) is used.  The variation of the action have to be
applied by independent variation of the fields
$g,\,\stackrel{(\pi)}{\omega}(x)$ (or equivalently
$\stackrel{(\pi)}{K}(x)$) and $\Psi(x),\,\overline{\Psi}(x)$. In
terms of the Euler-Lagrange variations, the least action
$\phantom{a}\delta S=0$ gives
\begin{equation}
\begin{array}{l}
\delta g^{\mu\nu}:\phantom{a} \stackrel{\circ}{G}_{\mu\nu}
+\frac{\delta (\sqrt{-g}\,L_{Q}^{(A)})}{\delta g_{\mu\nu}}=
-2{\ae}\,\frac{\delta (\sqrt{-g}\,L_{m}^{(\pi)})}{\delta
g^{\mu\nu}};\\
\delta\,\stackrel{(\pi)}{\omega}{}_{\nu}^{\phantom{j}\mu\rho}:\phantom{a}
\frac{\partial\,\stackrel{(A)}{\omega}{}_{\nu'}^{\phantom{j}\mu'\rho'}}{\partial
\stackrel{(\pi)}{\omega}{}_{\nu}^{\phantom{j}\mu\rho}}\,\frac{\delta}{\delta
\stackrel{(A)}{\omega}{}_{\nu'}^{\phantom{j}\mu'\rho'}}\,(\sqrt{-g}\,L_{Q}^{(A)})=-\frac{\delta
(\sqrt{-g}\,L_{m}^{(\pi)})}{\delta
\stackrel{(\pi)}{\omega}{}_{\nu}^{\phantom{j}\mu\rho}}; \\
\delta\,\Psi:\phantom{a} \frac{\delta\,
(\sqrt{-g}\,L_{m}^{(\pi)})}{\delta\,\Psi}=0; \quad
 \delta\,\overline{\Psi}:\phantom{a}\frac{\delta
\,(\sqrt{-g}\,L_{m}^{(\pi)})}{\delta \overline{\Psi}}=0,
\end{array}
\label{RU29}
\end{equation}
where $\stackrel{\circ}{G}_{\mu\nu}$ is the Einstein tensor
\begin{equation}
\begin{array}{l}
\stackrel{\circ}{G}_{\mu\nu}=\stackrel{\circ
}{R}_{\mu\nu}-\frac{1}{2}\,\stackrel{\circ}{R}\,g_{\mu\nu},
\end{array}
\label{PPP3}
\end{equation}
provided we have
\begin{equation}
\begin{array}{l}
\frac{1}{\sqrt{-g}}\,\frac{\delta (\sqrt{-g}\,L_{Q}^{(A)})}{\delta
g^{\mu\nu}}={\ae}\,\stackrel{(A)}{U}_{\mu\nu},
\end{array}
\label{RU30}
\end{equation}
where
\begin{equation}
\begin{array}{l}
\stackrel{(A)}{U}_{\mu\nu}=\frac{1}{{\ae}}\,[\stackrel{(A)}{K}{}^{\nu}_{\phantom{j}\mu\nu}\,
\stackrel{(A)}{K}{}^{\lambda}_{\phantom{l}\rho\lambda}-
\stackrel{(A)}{K}{}^{\lambda}_{\phantom{l}\mu\sigma}\,
\stackrel{(A)}{K}{}^{\sigma}_{\phantom{m}\rho\lambda}-
\frac{1}{2}\,g_{\mu\rho}\,(\stackrel{(A)}{K}{}^{\nu\sigma}_{\phantom{jm}\nu}\,
\stackrel{(A)}{K}{}^{\lambda}_{\phantom{l}\sigma\lambda}-
\stackrel{(A)}{K}{}^{\lambda\nu}_{\phantom{lj}\sigma}\,\stackrel{(A)}{K}{}^{\sigma}_{\phantom{m}\nu\lambda})],
\end{array}
\label{RP30}
\end{equation}
or
\begin{equation}
\begin{array}{l}
\stackrel{(A)}{U}_{\mu\nu}=\frac{1}{{\ae}}\,[-(\stackrel{(A)}{Q}{}^{\lambda}_{\phantom{l}\mu\nu}+
2\stackrel{(A)}{Q}{}_{(\mu\nu)}^{\phantom{(ij)}\lambda})(\stackrel{(A)}{Q}{}^{\nu}_{\phantom{j}\rho\lambda}+
2\stackrel{(A)}{Q}{}_{(\rho\lambda)}^{\phantom{(kl)}\nu})+4\stackrel{(A)}{Q}{}_{\mu}
\,\stackrel{(A)}{Q}{}_{\rho}+\frac{1}{2}\,g_{\mu\rho}(\stackrel{(A)}{Q}{}^{\sigma\nu\lambda}+\\
2\stackrel{(A)}{Q}{}^{(\nu\lambda)\sigma})
(\stackrel{(A)}{Q}{}_{\lambda\nu\sigma}+2\stackrel{(A)}{Q}{}_{(\nu\sigma)\lambda})-2g_{\mu\rho}\,\stackrel{(A)}{Q}{}^{\nu}\,
\stackrel{(A)}{Q}{}_{\nu}].
\end{array}
\label{RP31}
\end{equation}
By virtue of relation~(\ref{RU30}), we may recast the first equation
in~(\ref{RU29}) into the {\em first EC equation}
\begin{equation}
\begin{array}{l}
\stackrel{\circ}{G}_{\mu\nu}={\ae}\,\stackrel{(A)}{\theta}_{\mu\nu},
\end{array}
\label{RU35}
\end{equation}
where including the spin contributions directly into the
energy-momentum tensor, we introduce the canonical energy-momentum
tensor
\begin{equation}
\begin{array}{l}
\stackrel{(A)}{\theta}_{\mu\nu}:\phantom{a}=T_{\mu\nu}+
\stackrel{(A)}{U}_{\mu\nu}.
\end{array}
\label{RU34}
\end{equation}
For variations $\delta
\stackrel{(\pi)}{K}{}^{\nu}_{\phantom{\nu}\mu\rho}$ (or equivalently
$\stackrel{(\pi)}{\omega}{}^{\nu}_{\phantom{\nu}\mu\rho}$ ), the
$\phantom{a}\delta S=0$ gives the {\em second EC equation}
\begin{equation}
\begin{array}{l}
\frac{\partial\,\stackrel{(A)}{\omega}{}_{\nu'}^{\phantom{j}\mu'\rho'}}{\partial
\stackrel{(\pi)}{\omega}{}_{\nu}^{\phantom{j}\mu\rho}}(\stackrel{(A)}{\cal
T})\, \stackrel{(A)}{\cal T}{}_{\phantom{a}\mu'\rho'}^{\nu'}=
-\frac{1}{2}{\ae}\,
\stackrel{(\pi)}{S}{}_{\phantom{a}\mu\rho}^{\nu},
\end{array}
\label{RPP34}
\end{equation}
where the {\em modified torsion} reads
\begin{equation}
\begin{array}{l}
\stackrel{(A)}{\cal
T}{}_{\phantom{a}\mu\rho}^{\nu}:\phantom{a}=\frac{1}{2\sqrt{-g}}\,\frac{\delta
(\sqrt{-g}\,L_{Q}^{(A)})}{\delta
\stackrel{(A)}{\omega}{}_{\nu}^{\phantom{j}\mu\rho}}=
\stackrel{(A)}{Q}{}_{\phantom{a}\mu\rho}^{\nu}+\delta^{\nu}_{\mu}\,\stackrel{(A)}{Q}_{\rho}-
\delta^{\nu}_{\rho}\,\stackrel{(A)}{Q}_{\mu}.
\end{array}
\label{RPP}
\end{equation}
Thus, the equations of the standard EC  theory can be recovered for
$A=\pi$:
\begin{equation}
\begin{array}{l}
\stackrel{\circ
}{G}_{\mu\nu}={\ae}\,\stackrel{(\pi)}{\theta}_{\mu\nu},\quad
\stackrel{(\pi)}{\cal
T}{}_{\phantom{a}\mu\rho}^{\nu}=-\frac{1}{2}\,{\ae}\,
\stackrel{(\pi)}{S}{}_{\phantom{a}\mu\rho}^{\nu},
\end{array}
\label{EC1}
\end{equation}
in which the equation defining torsion is the algebraic type, such
that torsion at a given point in spacetime does not vanish only if
there is matter at this point, represented in the Lagrangian density
by a function which depends on torsion. Unlike the metric, which is
related to matter through a differential field equation, torsion
does not propagate. Combining (\ref{RP31}), (\ref{RPP}) and
(\ref{EC1})
 gives
\begin{equation}
\begin{array}{l}
\stackrel{(\pi)}{U}_{\mu\nu}=
{\ae}\left(-\stackrel{(\pi)}{S}{}^{\phantom{i}\,\rho}_{\mu\,[\lambda}\stackrel{(\pi)}{S}{}^{\lambda}_{\phantom{kl}\nu\rho]}-
\frac{1}{2}\stackrel{(\pi)}{S}{}^{\phantom{i}\rho\lambda}_{\mu}\stackrel{(\pi)}{S}_{\nu\rho\lambda}+\right.\\\left.
\frac{1}{4}\stackrel{(\pi)}{S}{}^{\rho\lambda}_{\phantom{ij}\mu}\stackrel{(\pi)}{S}{}_{\rho\lambda\nu}+
\frac{1}{8}g_{\mu\nu}(-4\stackrel{(\pi)}{S}{}^{\lambda}_{\phantom{l}\rho[\tau}\stackrel{(\pi)}{S}{}^{\rho\tau}_{\phantom{jm}\lambda]}+
\stackrel{(\pi)}{S}{}^{\rho\lambda\tau}\stackrel{(\pi)}{S}{}_{\rho\lambda\tau})\right).
\end{array}
\label{PS1}
\end{equation}
However, equations (\ref{EC1}) can be equivalently replaced by the
set of {\em modified EC equations} for $A=\sigma$:
\begin{equation}
\begin{array}{l}
\stackrel{\circ
}{G}_{\mu\nu}={\ae}\,\stackrel{(\sigma)}{\theta}_{\mu\nu},\quad
\Theta_{\nu'\mu\,\rho}^{\mu'\rho'\nu}(\stackrel{(\sigma)}{\cal T})\,
\stackrel{(\sigma)}{\cal
T}{}_{\phantom{a}\mu'\rho'}^{\nu'}=-\frac{1}{2}\,{\ae}\,
\stackrel{(\pi)}{S}{}_{\phantom{a}\mu\rho}^{\nu},
\end{array}
\label{EC222}
\end{equation}
where
\begin{equation}
\begin{array}{l}
\frac{\partial\,\stackrel{(\sigma)}{\omega}{}_{\nu'}^{\phantom{j}\mu'\rho'}}{\partial
\stackrel{(\pi)}{\omega}{}_{\nu}^{\phantom{j}\mu\rho}}(\stackrel{(\sigma)}{\cal
T}):\phantom{a}=
\Theta_{\nu'\mu\,\rho}^{\mu'\rho'\nu}(\stackrel{(\sigma)}{\cal
T})\equiv \Theta_{\nu'\mu\,\rho}^{\mu'\rho'\nu}(\pi(x),\,\sigma(x)),
\end{array}
\label{EC2}
\end{equation}
in which the torsion $\stackrel{(\sigma)}{\cal
T}{}_{\phantom{a}\mu\rho}^{\nu}$  is {\em dynamical} if only
$\Theta_{\nu'\mu\,\rho}^{\mu'\rho'\nu}(\pi(x),\,\sigma(x))\neq
\delta^{\mu'}_{\mu}\,\delta^{\rho'}_{\rho}\,\delta^{\nu}_{\nu'}$.
According to (\ref{RU34}), it is spin that generates a nonsymmetric
part in the canonical energy-momentum tensor and then, produces a
deviation from the Riemann geometry. The variation of
$S_{m}^{(\pi)}$ (\ref{RU26}) with respect to the metric-compatible
affine connection in the metric-affine variational formulation of
gravity is equivalent to the variation with respect to the torsion
(or contortion) tensor. Consequently, the dynamical spin density
$\stackrel{(\pi)}{s}{}_{ab}^{\phantom{ab}\mu}$ is identical with
\begin{equation}
\begin{array}{l}
\stackrel{(\pi)}{\Sigma}{}_{ab}^{\phantom{ab}\mu}=\frac{\partial
(\sqrt{-g}\,L_{m})^{(\pi)}}{\partial\Psi_{,\mu}}\,\stackrel{(\pi)}S_{ab}\,\Psi,
\end{array}
\label{P4}
\end{equation}
referred to as the {\em canonical spin density}. The canonical
tensor
$e\,\stackrel{(A)}{\theta}_{\mu\nu}=\stackrel{(A)}{\tau}_{\mu\nu}=\stackrel{(A)}{e}_{a
\nu}\,\stackrel{(A)}{\tau}{}^{\phantom{i}a}_{\mu}$ is generally not
symmetric, whereas the canonical energy-momentum density is
identical with the dynamical tetrad energy-momentum density
$e\stackrel{(A)}{\theta}{}^{\phantom{i}a}_{\mu}=\stackrel{(A)}{\tau}{}^{\phantom{i}a}_{\mu}$,
where $e:\phantom{a}=det|e^{\mu}_{a}|=\sqrt{-g}$. The relation
between the tetrad dynamical energy-momentum tensor and the metric
dynamical energy-momentum tensor for matter fields is
$\stackrel{(A)}{\theta}_{(\mu\nu)}=T_{\mu\nu}$. The
Belinfante-Rosenfeld relation, between the dynamical metric and
dynamical tetrad (canonical) energy-momentum tensors, can be written
as (see e.g. \cite{rf13}):
\begin{equation}
\begin{array}{l}
\stackrel{(A)}{\theta}_{\mu\nu}-T_{\mu\nu}=
\frac{1}{2}\,\nabla_{\nu}^\ast(\stackrel{(\pi)}{S}{}_{\mu\rho}^{\phantom{ik}\nu}-
\stackrel{(\pi)}{S}{}_{\rho\phantom{j}\mu}^{\phantom{k}\nu}+\stackrel{(\pi)}{S}{}^{\nu}_{\phantom{j}\mu\rho})=\stackrel{(A)}U_{\mu\nu},
\end{array}
\label{B6}
\end{equation}
where $ \nabla_{\mu}^\ast=\nabla_{\mu}-2\stackrel{(\pi)}{Q}_{\mu} $
is the {\em modified covariant derivative}. The conservation law for
the spin density results from antisymmetrizing the
Belinfante-Rosenfeld relation with respect to the indices
$\mu,\rho$:
\begin{equation}
\begin{array}{l}
\stackrel{(\pi)}{s}{}_{\mu\nu\phantom{k};\rho}^{\phantom{ij}\rho}=
\tau_{\mu\nu}-\tau_{\nu\mu}+2\stackrel{(\pi)}{Q}{}_{\rho}\,\stackrel{(\pi)}{s}{}_{\mu\nu}^{\phantom{ij}\rho}
\end{array}
\label{B8}
\end{equation}
or $ \frac{1}{2}\,\nabla_{\rho}^{\ast}\,
\stackrel{(\pi)}{S}{}_{\mu\nu}^{\phantom{ij}\rho}=\stackrel{(A)}{\theta}_{[\mu\nu]}$.

\subsection{TSSD-$U_{4}$ theory in the language of
differential forms} In  this subsection we re-derive the field
equations of the TSSD-$U_{4}$ theory by using the exterior calculus.
The fields  have to be expressed in terms of differential forms in
order to build the total Lagrangian 4-form as the appropriate
integrand of the action.  Let
$\stackrel{(A)}{\omega}{}^{ab}=\stackrel{(A)}{\omega}{}^{ab}_{\mu}\,\wedge\,d\,x^{\mu}$
be the 1-forms of corresponding connections assuming values in the
Lorentz Lie algebra. The action for gravitational field can be
written in the form
\begin{equation}
\begin{array}{l}
S_{g}=\stackrel{\circ}{S}+S_{Q}=-\frac{1}{4{\ae}}\,\int \star
\stackrel{\circ} {R}+S_{Q},
\end{array}
\label{RD3}
\end{equation}
where $\star$ denotes the Hodge dual. This is  a $C^\infty$-linear
map $\star:\Omega^p\to\Omega^{n-p}$, which acts on the wedge product
monomials of the basis 1-forms as $ \star(\vartheta^{a_1\cdots
a_p})=\varepsilon^{a_1\cdots a_n}e_{a_{p+1}\cdots{a_n}}. $ Here
$e_{a_i}\,\,(i=p+1,...,n)$ are understood as the down indexed
1-forms $e_{a_i}=o_{a_i b}\,\vartheta^b$ and
$\varepsilon^{a_1...a_n}$ is the total antisymmetric pseudo-tensor.
According to (\ref{R57C}), the relations between the Ricci scalars
read
\begin{equation}
\begin{array}{l}
\stackrel{\circ}{R}\equiv \stackrel{\circ\,(\sigma)}{R}_{cd}\,
\wedge\,\stackrel{\bullet}{\vartheta}{}^{c}\,
\wedge\,\stackrel{\bullet}{\vartheta}{}^{d}=\stackrel{\circ\,(\pi)}{R}_{cd}\,
\wedge\,\vartheta^{c}\, \wedge\,\vartheta^{d}.
\end{array}
\label{F7}
\end{equation}
Consider a phenomenological action of the spin-torsion interaction,
$S_{Q}$, such that the variation of the connection 1-form
$\stackrel{(A)}{\omega}{}^{ab}$ yields
\begin{equation}
\begin{array}{l}
\delta\,S_{Q}=\frac{1}{{\ae}}\,\int \star\stackrel{(A)}{\cal
T}_{ab}\, \wedge\,\delta \stackrel{(A)}{\omega}{}^{ab},
\end{array}
\label{RD4}
\end{equation}
where
\begin{equation}
\begin{array}{l}
 \star\stackrel{(A)}{\cal
T}_{ab}:\phantom{a}=\frac{1}{2}\,\star(\stackrel{(A)}{Q}_{a}\,\wedge\,\stackrel{(A)}{e}_{b})=
\stackrel{(A)}{Q}{}^{c}\,\wedge\,\stackrel{(A)}{\vartheta}{}^{d}\,\varepsilon_{cdab}=
\frac{1}{2}\,\stackrel{(A)}{Q}{}^{c}_{\phantom{a}
\mu\nu}\,\wedge\,\stackrel{(A)}{e}{}^{d}_{\phantom{a}\alpha
}\,\varepsilon_{abcd}\,\stackrel{(A)}{\vartheta}{}^{\mu\nu\alpha},
\end{array}
\label{RD5}
\end{equation}
here we used the abbreviated notations for the wedge product
monomials, $
\stackrel{(A)}{\vartheta}{}^{\mu\nu\alpha...}=\,\stackrel{(A)}{\vartheta}{}^{\mu}\,\wedge\,
\stackrel{(A)}{\vartheta}{}^{\nu}\,\wedge\,\stackrel{(A)}{\vartheta}{}^{\alpha}\,\wedge\,...$,
defined on the $U_{4}$ space, and that
\begin{equation}
\begin{array}{l}
\stackrel{(A)}{Q}{}^{a}=\stackrel{(A)}{D}\,\stackrel{(A)}{\vartheta}{}^{a}=d\,\stackrel{(A)}{\vartheta}{}^{a}+
\stackrel{(A)}{\omega}{}^{a}_{\phantom{a}
b}\,\wedge\,\stackrel{(A)}{\vartheta}{}^{b}.
\end{array}
\label{RD6}
\end{equation}
The variation of the action describing the macroscopic matter
sources $S_{m}^{(\pi)}$ with respect to the coframe $\vartheta^{a}$,
and connection 1-form $\stackrel{(\pi)}{\omega}{}^{ab}$ reads
\begin{equation}
\begin{array}{l}
\delta\,S_{m}=\int\delta\,L_{m}=\int \,(-\star
\stackrel{(A)}{\theta}{}_{a}\,\wedge\,\delta
\,\stackrel{(A)}{\vartheta}{}^{a}+\frac{1}{2}\,
\star\stackrel{(\pi)}{\Sigma}_{ab}\, \wedge\,\delta
\,\stackrel{(\pi)}{\omega}{}^{ab}),
\end{array}
\label{RD7}
\end{equation}
where $\star\stackrel{(A)}{\theta}_{a}$ is the dual 3-form relating
to the canonical energy-momentum tensor,
$\stackrel{(A)}{\theta}{}_{a}^{\mu}$, by
\begin{equation}
\begin{array}{l}
\star\,\stackrel{(A)}{\theta}_{a}=\frac{1}{3!}\,\stackrel{(A)}{\theta}{}_{a}^{\mu}\,\varepsilon_{\mu\nu\alpha\beta}
\,\stackrel{(A)}{\vartheta}{}^{\nu \alpha\beta}.
\end{array}
\label{RDD88}
\end{equation}
and
$\star\stackrel{(\pi)}{\Sigma}_{ab}=-\star\stackrel{(\pi)}{\Sigma}_{ba}$
is the dual 3-form corresponding to the canonical spin tensor, which
is identical with the dynamical spin tensor
$\stackrel{(\pi)}{S}_{abc}$, namely
\begin{equation}
\begin{array}{l}
\star\stackrel{(\pi)}{\Sigma}_{ab}=\stackrel{(\pi)}{S}{}^{\mu}_{\phantom{a}ab}\,\varepsilon_{\mu\nu\alpha\beta}
\,\stackrel{(\pi)}{\vartheta}{}^{\nu \alpha\beta}.
\end{array}
\label{RD8}
\end{equation}
The integral
\begin{equation}
\begin{array}{l}
\stackrel{\circ}{S}=-\frac{1}{4{\ae}}\,\int \star \stackrel{\circ}
{R}=-\frac{1}{4{\ae}}\,\int \star\stackrel{\circ\,(A)}{R}_{cd}\,
\wedge\,\stackrel{(A)}{\vartheta}{}^{c}\,
\wedge\,\stackrel{(A)}{\vartheta}{}^{d},
\end{array}
\label{F8}
\end{equation}
is the usual Einstein action, written in the language of the
exterior forms. Actually, writing explicitly the holonomic indices,
we have
\begin{equation}
\begin{array}{l}
\stackrel{\circ}{S}= -\frac{1}{8{\ae}}\,\int
\,\stackrel{\circ\,(A)}{R}{}_{\mu\nu}^{\phantom{ab}ab}\,\stackrel{(A)}{e}{}_{\alpha}^{\phantom{a}c}\,
\stackrel{(A)}{e}{}_{\beta}^{\phantom{a}d}\,\varepsilon_{abcd}\,\stackrel{(A)}{\vartheta}{}^{\mu\nu\alpha\beta}=
-\frac{1}{8{\ae}}\,\int
\,\stackrel{\circ\,(A)}{R}{}_{\mu\nu}^{\phantom{ab}ab}\,\varepsilon_{ab\alpha\beta}\,\varepsilon^{\mu\nu\alpha\beta}\,d\,\Omega.
\end{array}
\label{F9}
\end{equation}
Using the relations
\begin{equation}
\begin{array}{l}
\varepsilon_{ab\alpha\beta}\,\varepsilon^{\mu\nu\alpha\beta}=
-2e\,\stackrel{(A)}{e}{}_{ab}^{\mu\nu},\quad
\stackrel{(A)}{e}{}_{ab}^{\mu\nu}= \left|
  \begin{array}{ccc}
    \stackrel{(A)}{e}{}^{\mu}_{\phantom{a}a} & \stackrel{(A)}{e}{}^{\nu}_{\phantom{a}a}
    \\
    \stackrel{(A)}{e}{}^{\mu}_{\phantom{a}b} & \stackrel{(A)}{e}{}^{\nu}_{\phantom{a}b}
    \\
\end{array}
\right|,
\end{array}
\label{F10}
\end{equation}
we have
\begin{equation}
\begin{array}{l}
\stackrel{\circ}{S}= -\frac{1}{2{\ae}}\,\int
\,\stackrel{\circ\,(A)}{R}{}_{\mu\nu}^{\phantom{ab}ab}\,\stackrel{(A)}{e}{}_{\alpha}^{\phantom{a}[\mu}\,
\stackrel{(A)}{e}{}_{\beta}^{\phantom{a}\nu]}\,e\,d\,\Omega=
-\frac{1}{2{\ae}}\,\int \,\stackrel{\circ}{R}\,\sqrt{-g}\,d\,\Omega.
\end{array}
\label{F11}
\end{equation}
Also, one may readily verify that
\begin{equation}
\begin{array}{l}
\delta\,S_{Q}=\frac{1}{{\ae}}\,\int \stackrel{(A)}{\cal
T}{}_{\mu\nu}^{\phantom{ab}\beta}\, \delta
\stackrel{(A)}{\omega}{}_{\beta}^{\phantom{a}\mu\nu}.
\end{array}
\label{F12}
\end{equation}
Certainly,
\begin{equation}
\begin{array}{l}
\delta\,S_{Q}=\frac{1}{2{\ae}}\,\int
\stackrel{(A)}{Q}{}^{c}_{\phantom{a}\mu\nu}\,\stackrel{(A)}{e}{}^{\phantom{a}d}_{\alpha}\,\delta
\stackrel{(A)}{\omega}{}^{\phantom{a}ab}_{\beta}\,\varepsilon_{cdab}\,\stackrel{(A)}{\vartheta}{}^{\mu\nu\alpha\beta}=\\
-\frac{1}{2{\ae}}\,\int
\stackrel{(A)}{Q}{}^{c}_{\phantom{a}\mu\nu}\,\delta
\stackrel{(A)}{\omega}{}^{\phantom{a}ab}_{\beta}\,\varepsilon_{\alpha
cab}\varepsilon^{\alpha\mu\nu\beta}\,d\,\Omega.
\end{array}
\label{F13}
\end{equation}
Using the relations $\varepsilon_{\alpha
cab}\,\varepsilon^{\alpha\mu\nu\beta}=
-e\,\stackrel{(A)}{e}{}_{cab}^{\mu\nu\beta}$, where
\begin{equation}
\begin{array}{l}
\stackrel{(A)}{e}{}_{cab}^{\mu\nu\beta}= \left|
  \begin{array}{ccc}
    \stackrel{(A)}{e}{}^{\mu}_{\phantom{a}c} & \stackrel{(A)}{e}{}^{\nu}_{\phantom{a}c}& \stackrel{(A)}{e}{}^{\beta}_{\phantom{a}c}
    \\
    \stackrel{(A)}{e}{}^{\mu}_{\phantom{a}a} & \stackrel{(A)}{e}{}^{\nu}_{\phantom{a}a}& \stackrel{(A)}{e}{}^{\beta}_{\phantom{a}a}
    \\
    \stackrel{(A)}{e}{}^{\mu}_{\phantom{a}b} & \stackrel{(A)}{e}{}^{\nu}_{\phantom{a}b}& \stackrel{(A)}{e}{}^{\beta}_{\phantom{a}b}
    \\
\end{array}
\right|,
\end{array}
\label{F10}
\end{equation}
we obtain
\begin{equation}
\begin{array}{l}
\frac{1}{2}\, \stackrel{(A)}{Q}{}^{c}_{\phantom{a}\mu\nu}\,
\stackrel{(A)}{e}{}_{cab}^{\mu\nu\beta}=
\stackrel{(A)}{e}{}_{a}^{\mu}\,\stackrel{(A)}{e}{}_{b}^{\nu}\stackrel{(A)}{\cal
T}{}_{\mu\nu}^{\phantom{ab}\beta}.
\end{array}
\label{F14}
\end{equation}
and (\ref{F13}) gives (\ref{F12}). The variation of the total
action, given by the sum of the gravitational field action and the
matter action, with respect to the $e^{a}$,
$\stackrel{(\pi)}{\omega}{}^{ab}$, and $\Psi$, gives
\begin{equation}
\begin{array}{l}
1)\phantom{a}\frac{1}{2}\,\stackrel{\circ\,(A)}{R}_{cd}\,
\wedge\,\stackrel{(A)}{\vartheta}{}^{c}={\ae}\,\stackrel{(A)}{\theta}_{d},\quad
2)\phantom{a}\frac{\partial\,\stackrel{(A)}{\omega}{}^{a'b'}}{\partial\,\stackrel{(\pi)}{\omega}{}^{ab}}\,\wedge\,
\star\stackrel{(A)}{\cal T}_{a'b'}=
-\frac{1}{2}\,{\ae}\,\star\stackrel{(\pi)}{\Sigma}_{ab},\\
3)\phantom{a}\frac{\delta \,L_{m}^{(\pi)}}{\delta \Psi}=0,\quad
\frac{\delta \,L_{m}^{(\pi)}}{\delta \overline{\Psi}}=0,
\end{array}
\label{RD15}
\end{equation}
In the tensor language then  the first equation in (\ref{RD15})
coincides with the tensorial equation (\ref{RU35}). To prove this,
we may recast it into the form
\begin{equation}
\begin{array}{l}
\frac{1}{4}\,\stackrel{\circ\,(A)}{R}_{\mu\nu}^{\phantom{ab}ab}\,
\stackrel{(A)}{\vartheta}{}^{c}_{\alpha}\,\varepsilon_{abcd}\,\varepsilon^{\mu\nu\alpha\beta}=
\frac{1}{3!}\,{\ae}\stackrel{(A)}{\theta}{}_{d}^{a}\,\varepsilon_{a\mu\nu\alpha}\,\varepsilon^{\mu\nu\alpha\beta}
\,\stackrel{(A)}{\vartheta}{}^{\nu \alpha\beta},
\end{array}
\label{RD16}
\end{equation}
such that
\begin{equation}
\begin{array}{l}
-\frac{e}{4}\,\stackrel{\circ\,(A)}{R}_{\mu\nu}^{\phantom{ab}ab}\,
\stackrel{(A)}{e}{}^{\mu\nu\beta}_{abc}=
{\ae}\stackrel{(A)}{\theta}{}_{c}^{a}\,e\,\stackrel{(A)}{e}{}^{\phantom{a}\beta}_{a}.
\end{array}
\label{RD17}
\end{equation}
Making use of the relation
\begin{equation}
\begin{array}{l}
-\stackrel{\circ\,(A)}{R}_{\mu\nu}^{\phantom{ab}ab}\,
\stackrel{(A)}{e}{}^{\mu\nu\beta}_{abc}=-2\stackrel{\circ\,(A)}{R}\,\stackrel{(A)}{e}{}^{\phantom{a}\beta}_{c}+
4\stackrel{\circ\,(A)}{R}{}^{\phantom{a}\beta}_{c},
\end{array}
\label{F18}
\end{equation}
finally, gives
\begin{equation}
\begin{array}{l}
\stackrel{\circ\,(A)}{R}{}^{\phantom{a}\beta}_{c}-\frac{1}{2}\,\stackrel{(A)}{e}{}^{\phantom{a}\beta}_{c}\,\stackrel{\circ}{R}=
{\ae}\stackrel{(A)}{\theta}{}_{\alpha}^{\beta}.
\end{array}
\label{F19}
\end{equation}
We may evaluate the second equation in (\ref{RD15}) as
\begin{equation}
\begin{array}{l}
\frac{\partial\,\stackrel{(A)}{\omega}{}^{cd}}{\partial\,\stackrel{(\pi)}{\omega}{}^{c'd'}}\,\wedge\,
\frac{1}{2}\,\stackrel{(A)}{Q}{}^{a}\,\wedge\,\stackrel{(A)}{e}{}^{b}\,\varepsilon_{abcd}
=- \frac{1}{2}\,{\ae}\,\star\stackrel{(\pi)}{\Sigma}_{cd},
\end{array}
\label{F20}
\end{equation}
and that
\begin{equation}
\begin{array}{l}
\frac{\partial\,\stackrel{(A)}{\omega}{}^{cd}_{\beta}}{\partial\,\stackrel{(\pi)}{\omega}{}^{c'd'}_{\beta'}}\,
\frac{1}{2}\,\stackrel{(A)}{Q}{}^{a}_{\phantom{a}\mu\nu}\,
\stackrel{(A)}{e}{}^{\phantom{a}b}_{\alpha}\,\varepsilon_{abcd}\,\varepsilon^{\mu\nu\alpha\beta}=
-
\frac{1}{2}\,{\ae}\,\stackrel{(\pi)}{S}{}^{a}_{\phantom{a}c'd'}\,\varepsilon_{a\mu\nu\alpha}\,\varepsilon^{\mu\nu\alpha\beta'},
\end{array}
\label{F21}
\end{equation}
so
\begin{equation}
\begin{array}{l}
\frac{\partial\,\stackrel{(A)}{\omega}{}^{cd}_{\beta}}{\partial\,\stackrel{(\pi)}{\omega}{}^{c'd'}_{\beta'}}\,
\frac{e}{2}\,\stackrel{(A)}{Q}{}^{a}_{\phantom{a}\mu\nu}\,\stackrel{(A)}{e}{}_{acd}^{\mu\nu\beta}=
-
\frac{e}{2}\,{\ae}\,\stackrel{(\pi)}{S}{}_{\phantom{a}c'd'}^{a}\,\stackrel{(A)}{e}{}_{a}^{\beta'}=
-
\frac{e}{2}\,{\ae}\,\stackrel{(\pi)}{S}{}_{\phantom{a}c'd'}^{\beta'}.
\end{array}
\label{F22}
\end{equation}
Taking into account relation (\ref{F14}), we then obtain
\begin{equation}
\begin{array}{l}
\frac{\partial\,\stackrel{(A)}{\omega}{}^{cd}_{\beta}}{\partial\,\stackrel{(\pi)}{\omega}{}^{c'd'}_{\beta'}}\,
\stackrel{(A)}{\cal T}{}_{cd}^{\phantom{ab}\beta}=-
\frac{1}{2}\,{\ae}\,\stackrel{(\pi)}{S}{}_{\phantom{a}c'd'}^{\beta'},
\end{array}
\label{F24}
\end{equation}
which concises with the equation (\ref{RPP34}).

\subsection{Short-range spin-spin interaction}
In this subsection we derive the equations of the short-range
propagating torsion, which is of fundamental importance from a view
point of microphysics. This, together with the torsion waves, may
contribute a new special polarized effect in the current experiments
of a verification of gravitational spin-torsion interaction. These
experiments include neutron interferometry, neutron spin rotation
induced by torsion in vacuum, anomalous spin-dependent forces with a
polarized mass torsion pendulum, space-based searches for spin in
gravity, the neutrino oscillations, etc., see e.g. \cite{rf24}. For
instance, in the case of torsion, the fact that neutrino
oscillations are possible also if neutrinos are massless is very
important because, in general, it is thought that if one finds
neutrino oscillations the neutrinos must have a mass different from
$0$. This would be an interesting topic not discussed in this paper.
A remarkable feature of the present theoretical work is describing a
propagating torsion (\ref{EC222}). Furthermore, this in a natural
way can be made a short-range propagating torsion. Actually, from
the equations (\ref{RPP34}) and (\ref{RPP}), we see that it is the
spin $\stackrel{(\pi)}{S}$ and spacetime deformations $\pi(x)$ and
$\sigma(x)$ that define the torsion $\stackrel{(A)}{Q}$:
\begin{equation}
\begin{array}{l}
\stackrel{(A)}{Q}{}_{\phantom{a}\mu\rho}^{\nu}=\stackrel{(A)}{\cal
T}{}_{\phantom{a}\mu\rho}^{\nu}+
\frac{1}{2}\delta^{\nu}_{\mu}\,\stackrel{(A)}{\cal
T}{}_{\phantom{a}\rho\lambda}^{\lambda}-
\frac{1}{2}\delta^{\nu}_{\rho}\,\stackrel{(A)}{\cal
T}{}_{\phantom{a}\mu\lambda}^{\lambda},
\end{array}
\label{Q}
\end{equation}
which through definitions (\ref{RP31}), (\ref{RU34}) and the field
equation(\ref{RU35}), in turn, defines Einstein's field tensor
$\stackrel{\circ}{G}$. A generic spacetime deformation, $\pi(x)$,
consists of two ingredient deformations
($\stackrel{\bullet}{\pi}(\stackrel{\bullet}{x})$, \,$\sigma(x)$) of
the orthonormal frame. Whereas, when the deformation matrix
$\stackrel{\bullet}{\pi}(\stackrel{\bullet}{x})$ implies a peculiar
condition (\ref{R49}), the choice of the $\sigma(x)$ is not fixed
yet. This allows us to impose a physical constraint upon the
spacetime deformation $\sigma(x)$:
\begin{equation}
\begin{array}{l}
\Theta_{\nu'\mu\,\rho}^{\mu'\rho'\nu}(\pi(x),\,\sigma(x))=
(\square+M^{2}_{{\cal T}})\stackrel{(\sigma)}{\cal
T}{}_{\phantom{a}\mu\rho}^{\nu}\,(\stackrel{(\sigma)}{\cal
T}{}^{-1}){}^{\phantom{a}\mu'\rho'}_{\nu'},
\end{array}
\label{EC333}
\end{equation}
where $\square$  is a generalization of the d'Alembertian operator
for covariant derivatives defined on the RC manifold, $U_{4}$. Then,
the set of {\em modified EC equations} (\ref{EC222}) reduces to
\begin{equation}
\begin{array}{l}
\stackrel{\circ
}{G}_{\mu\nu}={\ae}\,\stackrel{(\sigma)}{\theta}_{\mu\nu},\quad
(\square+M^{2}_{{\cal T}})\stackrel{(\sigma)}{\cal
T}{}_{\phantom{a}\mu\rho}^{\nu}=-\frac{1}{2}\,{\ae}\,
\stackrel{(\pi)}{S}{}_{\phantom{a}\mu\rho}^{\nu},
\end{array}
\label{EC444}
\end{equation}
which describe the short-range propagating torsion and spin-spin
interaction. Actually, at large distances  $r >\lambda_{{\cal
T}}\equiv \frac{\hbar}{M_{\cal T}\,c} $ (Compton length), torsion
vanishes $\stackrel{(\sigma)}{\cal T}(r)=0$. To carry through this
theory in full generality, for example, we may explicitly write the
torsionic equation for the Dirac spinor matter source coupled to the
metric and to the torsion. Both are contained implicitly in the
connection $ \stackrel{(\pi)}{\omega}{}^{ba}_{\phantom{aa} \mu} $.
The Dirac spinor field defined in the TSSD-$U_{4}$ theory coincides
with a conventional formalism of the spinor field defined on the RC
spacetime \cite{rf9,rf13}.  The Lagrangian of spinor field written
for any frame of reference is
\begin{equation}
\begin{array}{l}

e\,L_\psi^{(\pi)}=\frac{ie}{2}\,(\bar{\psi}\,\stackrel{(\pi)}{g}{}^{\mu}\psi_{,\,\mu}-
\bar{\psi}_{,\,\mu}\,\stackrel{(\pi)}{g}{}^{\mu}\psi)-\frac{ie}{2}\,\bar{\psi}\,\{\stackrel{(\pi)}{g}{}^{\mu}\,
\,\stackrel{(\pi)}{\Gamma}_{\mu}\}\,\psi_{,\,\mu} -
m\,e\,\bar{\psi}\psi,
\end{array}
\label{SS2}
\end{equation}
where $\gamma^{a}$ are Dirac matrices, and
$\stackrel{(A)}{g}{}^{\mu}=\stackrel{(A)}{e}{}_{a}^{\phantom{a}\mu}\,\gamma^{a}$.
The spinor connection $\stackrel{(\pi)}{\Gamma}_{\mu}$ is given, up
to the addition of an arbitrary vector multiple of $I$, by the {\em
Fock-Ivanenko coefficients}
\begin{equation}
\begin{array}{l}
\stackrel{(\pi)}{\Gamma}_\mu=-\frac{1}{4}\,\stackrel{(\pi)}{\omega}{}_{ab
\mu}\,\gamma^{a}\,
\gamma^{b}=-\frac{1}{2}\,\stackrel{(\pi)}{\omega}{}_{ab
\mu}\,S^{ab}=
-\frac{1}{8}\,e^{\nu}_{c;\,\mu}[\stackrel{(\pi)}{g}_{\nu},\,\gamma^{c}]=
\frac{1}{8}\,[\stackrel{(\pi)}{g}{}^{\nu}_{\phantom{j};\,\mu},\,\stackrel{(\pi)}{g}_{\nu}],
\end{array}
\label{S3}
\end{equation}
with $S^{ab}=\frac{1}{2}\,\gamma^{[a}\,\gamma^{b]}=
\frac{1}{4}\,(\gamma^{a}\,\gamma^{b}-\gamma^{b}\,\gamma^{a})$ - the
spinor representation. Therefore, in the absence of other sources of
torsion, the RC manifold, $U_{4}$, with a Dirac field will be
characterized by the Lagrangian
\begin{equation}
\begin{array}{l}

e\,L_\psi^{(\pi)}=\frac{ie}{2}\,(\bar{\psi}\,\stackrel{(\pi)}{g}{}^{\mu}\psi_{,\,\mu}-
\bar{\psi}_{,\,\mu}\,\stackrel{(\pi)}{g}{}^{\mu}\psi)+\frac{ie}{8}\,
\stackrel{(\pi)}{\omega}_{ab
\mu}\,\bar{\psi}\{\stackrel{(\pi)}{g}{}^{\mu},\,\gamma^{a}
\,\gamma^{b}\} \,\psi_{,\,\mu} - m\,e\,\bar{\psi}\psi,
\end{array}
\label{S4}
\end{equation}
Using the identity
$\{\stackrel{(\pi)}{g}{}^{\mu},\,\stackrel{(\pi)}{g}{}^{\nu}\,
\stackrel{(\pi)}{g}{}^{\rho}\}=2\,\stackrel{(\pi)}{g}{}^{[\mu}\,\stackrel{(\pi)}{g}{}^{\nu}\,\stackrel{(\pi)}{g}{}^{\rho]}$,
the totally antisymmetric spin corresponding to the Lagrangian
density (\ref{S4}) is
\begin{equation}
\begin{array}{l}
\stackrel{(\pi)}{S}{}^{\mu\nu\rho}=\stackrel{(\pi)}{S}{}^{[\mu\nu\rho]}=
\frac{i}{2}\,\bar{\psi}\,\stackrel{(\pi)}{g}{}^{[\mu}
\stackrel{(\pi)}{g}{}^{\nu} \stackrel{(\pi)}{g}{}^{\rho]}\,\psi.
\end{array}
\label{S6}
\end{equation}
Consequently,  we may recast the torsionic equation into the form
\begin{equation}
\begin{array}{l}
(\square+M^{2}_{{\cal T}})\stackrel{(\sigma)}{\cal
T}{}_{\phantom{a}\mu\rho}^{\nu}=-\frac{i}{4}\,{\ae}\,
\bar{\psi}\,\stackrel{(\pi)}{g}{}^{[\mu} \stackrel{(\pi)}{g}{}^{\nu}
\stackrel{(\pi)}{g}{}^{\rho]}\,\psi,
\end{array}
\label{EC555}
\end{equation}
where $\psi$ implies the {\em Heisenberg-Ivanenko} nonlinear
equation which can be derived from the Lagrangian (\ref{S4}) by
means of the standard calculation:
\begin{equation}
\begin{array}{l}
i\,\stackrel{(\pi)}{g}{}^{\rho}\,\psi_{:\,\rho}-
\frac{3{\ae}}{8}(\bar{\psi}\,\stackrel{(\pi)}{g}{}_{\rho}\gamma^5\psi)\stackrel{(\pi)}{g}{}^{\rho}\gamma^5\,\psi=m\psi,
\end{array}
\label{S16}
\end{equation}
This is the Dirac equation written in the so-called second-order
formalism, in which the contortion tensor is given explicitly in
terms of the spin sources. In the limit when we neglect the usual
Riemannian terms depending on the metric and the curvature
($\partial_{;\,\mu}\rightarrow
\partial_{\mu} $), as we are interested only in the spin-torsion
interaction, we then have
\begin{equation}
\begin{array}{l}
\stackrel{(\sigma)}{\cal
T}{}_{\phantom{a}\mu\rho}^{\nu}(x)=\frac{{\ae}}{2}\int
G_{F}(x,\,x')\,\stackrel{(\pi)}{S}{}_{\phantom{a}\mu\rho}^{\nu}(x')\,d^{4}x'
\end{array}
\label{SH1}
\end{equation}
where the  Feynman propagator reads
\begin{equation}
\begin{array}{l}
G_{F}(x,\,x')=\left(
                \begin{array}{c}
                  -\frac{1}{4\pi}\delta(s)+\frac{M_{\cal
T}}{8\pi\sqrt{s}}H^{(1)}_{1}(M_{\cal T}\sqrt{s})\quad\mbox{if}\quad
s\geq 0 \\\\
                  -\frac{iM_{\cal
T}}{4\pi^{2}\sqrt{-s}}K_{1}(M_{\cal T}\sqrt{-s})\quad\mbox{if}\quad
s< 0, \\
                \end{array}
              \right)

\end{array}
\label{SH2}
\end{equation}
provided $s=(x-x')^{2}$,  $\, H^{(1)}_{1}$ is the Hankel function of
first kind and $K_{1}$ is a modified Bessel function. A detailed
analysis and calculations on the more general MAG theory with
dynamical torsion in context of TSSD formulation of post-Riemannian
geometry will be presented in another paper to follow at a later
date.

\section{Concluding remarks}
We show how the curvature and torsion, which are properties of a
connection of geometry under consideration, will come into being?
The theoretical significance resides in constructing the theory of
TSSD as a guiding principle. In this, we have to separate from the
very outset the case of teleparallel gravity, in which the torsion
is a propagating field, from the case of Einstein-Cartan (EC)
theory, in which it is not. This motivates our specific choice of
the general spacetime deformation $\pi(x)$ of the orthonormal frame,
to be consisted of two ingredient deformations
($\stackrel{\bullet}{\pi}(\stackrel{\bullet}{x})$, \,$\sigma(x)$).
We choose the first deformation matrix
$\stackrel{\bullet}{\pi}(\stackrel{\bullet}{x})$ in such a way
((\ref{R49}) or (\ref{R54})) that the deformed connection is set as
the Weitzenb\"ock connection. We consider then the different affine
connections (\ref{E13}), with different curvature and torsion,
associated with the deformation-related frame connection
(\ref{RR64}) and a general spin connection (\ref{RR65}). They are
independent of tetrad fields and their derivatives. Therefore, we
separate the notions of space and connections, namely we take a
spacetime simply as a manifold, and affine connections as additional
structures- the metric-affine formulation of gravity. Defining a
translation in the connection space, we construct the TSSD-versions
of the theory of teleparallel gravity and the EC theory. It is
remarkable that the equations of the standard EC theory, in which
the equation defining torsion is the algebraic type and, in fact, no
propagation of torsion is allowed, can be equivalently replaced by
the set of {\em modified EC equations} in which the torsion, in
general, is dynamical. Furthermore, we assume a special physical
ansatz (\ref{EC333}) for the spacetime deformation $\sigma(x)$,
which yields the short-range propagating spin-spin interaction.
This, together with the torsion waves, may contribute a new special
polarized effect in the current experiments of a verification of
gravitational spin-torsion interaction.

\section*{Acknowledgments}
Helpful and knowledgable comments and suggestions from the anonymous
referees are much appreciated.

\end{document}